\documentclass[
 ,twocolumn%
 ,secnumarabic%
,amssymb, amsmath,nobibnotes, aps]{revtex4-1}

\usepackage[table]{xcolor}

\usepackage{bm}
\usepackage{color}

\usepackage{comment}
\usepackage{extraarrows}
\usepackage{graphicx}
\usepackage{dcolumn}
\usepackage{bm}

\usepackage{mathrsfs}  

\def\bal#1\eal{\begin{align}#1\end{align}}
\def\mbf#1{\mbox{\boldmath $#1$}}

\usepackage{amssymb,amsfonts,amsmath}

\makeatletter 
\def\@cite#1{(#1)}
\makeatother

\def\j{{j^*}}

\def\a{{a^*}}
\def\J{{J^*}}

\def\A{{\bf A}}
\def\m{{\mathfrak m}}
\def\r{{\mathfrak r}}
\def\G{\Gamma}

\definecolor{LG}{gray}{0.55}
\definecolor{LLG}{gray}{0.75}
\definecolor{LLLG}{gray}{0.9}

\begin{document}

\newcommand{\tabenv}[1][\linewidth]{\def\@captype{table}}

\title{Sensitivity and Network Topology in Chemical Reaction Systems}

\author{Takashi Okada$^1$  and Atsushi Mochizuki$^{1,2}$}
\affiliation{%
$^1$Theoretical Biology Laboratory, RIKEN, Wako 351-0198, Japan \\
$^2$CREST, JST
4-1-8 Honcho, Kawaguchi 332-0012, Japan
}



\begin{abstract}

In living cells,  biochemical reactions are catalyzed by specific enzymes and connect to one another by sharing substrates and products, forming complex  networks.   In our previous studies, we established a framework determining the responses  to enzyme perturbations  only from network topology, and then proved a theorem, called the law of localization, explaining response patterns in terms of network topology. In this paper, we generalize these results to reaction networks with conserved concentrations, which allows us to study any reaction systems. We also propose novel network characteristics quantifying  robustness. We  compare E. coli metabolic network with randomly rewired networks, and find that the robustness of the E. coli network is significantly higher than that of the random networks.

\begin{description}
\item[PACS numbers]
82.20.-w, 87.10.Ed, 87.18.-h
\end{description}
\end{abstract}

   
\maketitle

\section{Introduction}
In living cells, there are  many  chemical reactions, and they forms complex networks such as metabolic networks and signal transduction networks. The dynamics of concentration of chemicals resulting from these networks is considered to be the origin of  physiological functions. Today,  a huge information about reaction networks is  available in databases  such as KEGG \cite{KEGG}, Reactome \cite{Reactome} and BioCyc \cite{BioCyc}.   

\begin{figure}[h] 
  \includegraphics[width=8.0cm,bb=0 10 300 75]{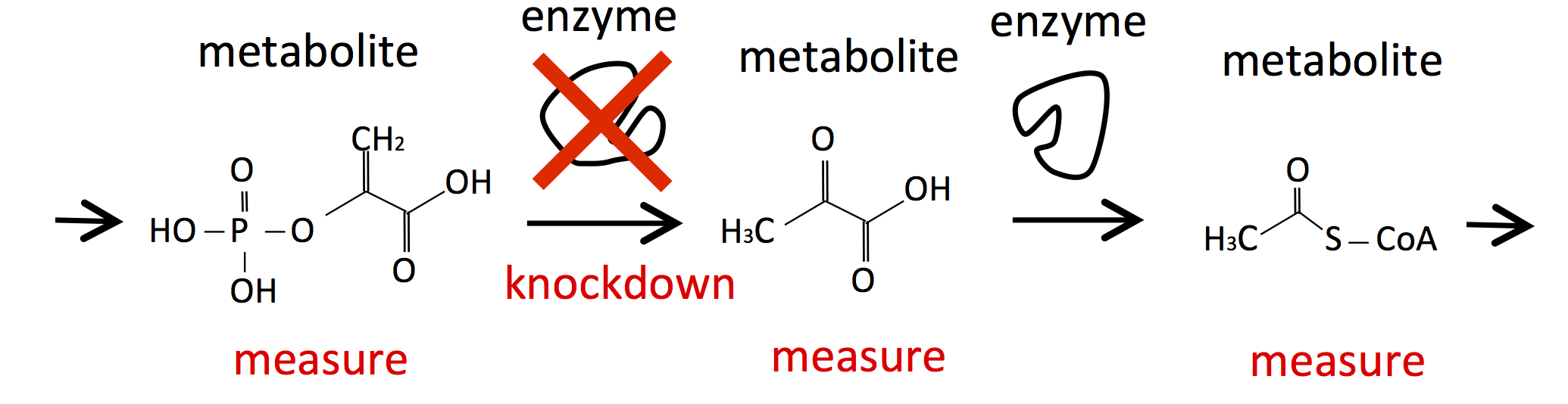}
   \caption{ Sensitivity analysis. The concentration changes of metabolites under the decrease of the amount/activity of an  enzyme protein are measured.}
   \label{fig:ko}
\end{figure}

Dynamics and steady state of reaction networks are determined by various factors in systems; reaction-specific enzymes, input signals from outside of systems, and initial conditions in some cases.  One of the standard approaches to elucidate the dynamics is a sensitivity experiment;  for example, in the research of metabolism,  changes in concentrations of metabolites induced by perturbation in amounts/activities of enzymes are examined \cite{Ishii} (see Fig \ref{fig:ko}).  

In previous theoretical studies \cite{Mochizuki, monomolecular}, under an implicit assumption that the system has no conserved concentrations, one of the authors and his collaborator  showed that the sensitivity of steady state to reaction rate parameters, which correspond to enzyme activities/amounts,  is  determined only from the structure of the chemical reaction network ({\it structural sensitivity analysis}). Such a structural approach has a great advantage because it is almost impossible to measure the precise kinetics and  parameters of chemical reactions in living cells. We then proved a novel  theorem, called the {\it law of localization} \cite{OM}. The law of localization  characterizes  subnetworks by non-positive indices (see $\lambda(\Gamma)$ in \eqref{LoL}) and identifies those which confine the effect of perturbations inside them. We call such subnetworks {\it buffering structures}.

One of the main purposes of this paper is to extend  our previous method to   reaction systems where  some  of concentrations are conserved during the dynamics. 
 For example, in the MAPK  pathway,   while the ratio between activated and inactivated kinase concentrations changes after stimuli, the total number  of them is  conserved in time. In the presence of such conserved quantities, steady state concentrations and fluxes are influenced not only by reaction rate parameters but also by the initial values of conserved quantities. We take  into account constraints coming from conserved quantities in the sensitivity analysis. 


Another purpose of this paper is to explore biological meanings of buffering structures.  Buffering structures provide the system with robustness to enzymatic fluctuations since they confine the effects inside them. 
Thus, we expect that possessing buffering structures are advantageous for living systems and networks with more buffering structures are selected during evolutionary process.  In order to refine these expectations,  we quantify  robustness  of reaction networks and compare  robustness of  Escherichia coli  metabolism with artificial random networks.  

The rest of the paper is organized as follows: In Sec. II, we generalize the method of structural sensitivity analysis. In Sec. III, we illustrate the method in a simple reaction network. In Sec. IV, we extend  the law of localization to any reaction  networks. In Sec. V, as  applications, we study two signal transduction networks. The first network has conserved quantities and also include regulations from non-substrate chemicals. The second network also has conserved quantities. In Sec. VI, we propose  network characteristics for robustness, and compare the E. coli network and random networks. The detailed explanation of the formulation and the proof of the law of localization are written in Appendix.

\section{Structural Sensitivity Analysis}\label{sec:Methods}
The structural sensitivity analysis  is a systematic method of determining sensitivity of steady states to rate parameter perturbations from reaction network information alone \cite{Mochizuki, monomolecular}. 
We generalize the structural sensitivity analysis to networks with conserved concentrations. In the generalized formulation,  not only the sensitivity to rate parameter perturbations but also that to   initial conditions on conserved concentrations are determined from network information. 

We label  chemical species by  $m \,(m=1,\ldots,M)$ and reactions by  $j\,(j=1,\ldots,R)$. 
In general, a macroscopic state of a spatially homogeneous chemical reaction system is specified by the concentrations $x_m(t)$ and obeys the following differential equations \cite{Mochizuki, monomolecular,F1,F2,OM}
\begin{align}
\frac{d x_m}{dt}= \sum_{j=1}^R \nu_{mj} r_j(k_j;x) \label{ode}.
\end{align}
Here, the $M\times R$ matrix $\nu$ is called the stoichiometric matrix:  If the stoichiometry of the reaction $j$ among   molecules $X_m$ is 
\begin{align}
j: \ \sum_{m=1}^M y^j_m X_m \rightarrow \sum_{m=1}^M \bar{y}^j_m X_m \ \ \\
 (X_m{:\ \rm types\ of\ molecules}),\nonumber
\end{align}
then the component $\nu_{mj}$ is defined as
\begin{align}
\nu_{mj}\equiv  \bar{y}^j_m -y^j_m.
\end{align}

The reaction rate function $r_j$ is called a flux, which depends on the  chemical concentrations $x$ and 
also on a reaction rate parameter $k_j$. 
We do not assume specific forms for the flux functions except that each flux  is an increasing function of its substrate concentration;
\begin{align}
\frac{\partial r_j}{\partial x_m} > 0 \  {\rm if} \ y^j_m >0, \ {\rm \ otherwise \ } \ \frac{\partial r_j}{\partial x_m} =0.\label{noreg}
\end{align}
Below, we abbreviate $\frac{\partial r_j}{\partial x_m}$ evaluated at steady state  as $r_{jm}$. Usual kinetics, such as the mass-action  and the Michaelis-Menten kinetics, satisfies this condition.  

In the context of metabolic reaction systems,  $x_m$ is the concentration of the $m$-th metabolite, and the $j$-th reaction rate parameter $k_j$ corresponds  the enzyme activity/amount catalyzing the $j$-th reaction.

We introduce notations about the kernel (right-null) and cokernel (left-null) spaces of $\nu$. We choose bases of the kernel and cokernel spaces, ${\rm ker}\ \nu$ and ${\rm coker} \ \nu$, as $\{{\mbf c}^\alpha\}_{(\alpha =1,\ldots,N)} $and $\{{\mbf d}_a\}_{(a=1,\ldots,N_c)}$, where  $N$ and $N_c$ denote their dimensions.
The kernel space of $\nu$  corresponds to steady state fluxes \cite{Palsson, FBA1,FBA2,FBA3}. Namely, steady state fluxes satisfy
\bal
r_j =\sum_{\alpha=1}^N \mu^\alpha  c^{\alpha}_j \label{kernel}
\eal
where  $c^\alpha_j$ is the $j$-th component of the vector ${\mbf c}^\alpha$, and $\mu^\alpha$ are  coefficients.
On the other hand, the cokernel space is related with conserved quantities. For every $a\in\{1,2,\ldots, N_c\}$, 
 \bal
l_a\equiv {\mbf d_a} \cdot {\mbf x } 
 \eal
 is constant in time, where $\mbf x$ is the vector $(x_1,\ldots,x_M)$. $l_a$ is a constant associated with ${\mbf d}_a$.
 
 In the previous papers \cite{Mochizuki,monomolecular,OM}, the condition $N_c =0$ is assumed, and so steady state concentrations and fluxes are functions of rate parameters $k_i$. However, when $ N_c >0$,  steady state also depends  on  initial conditions on conserved concentrations, i.e. $\{l_a\}$.  Therefore, in this case, there are two types of perturbations; the perturbation of the rate parameter, $k_\j\rightarrow k_\j+\delta k_\j$ and that of the $\a$-th conserved quantity, $l_\a \rightarrow l_\a + \delta l_\a$, where $j=\j$ and $a=\a$ refer to the perturbed rate parameter and conserved quantity respectively. 

To  treat two types of perturbations in a unified way, we introduce generalized parameters $K_J\, (J=1,\ldots,R+N_c)$ as
\bal
\{ K_1,\ldots,K_R, K_{R+1},\ldots, K_{R+N_c}\}\nonumber \\ \equiv \{k_1,\ldots, k_R,l_1,\ldots,l_{N_c}\}.
\eal
We determine the concentration changes $\delta_\J x_m$ and the flux changes $\delta_\J r_j$ at steady state under the $\J$-th perturbation, $K_\J\rightarrow K_\J +\delta K_\J$ $(\J=1,\ldots,R+N_c)$. 

As shown in Appendix, responses to all perturbations are obtained simultaneously from the following matrix equation;
\begin{align}
{\bf A} \left(
\begin{array}{cccc}
\delta_1 {  \mbf x} &  \ldots  &\delta_{R+N_c} { \mbf  x}  \\\hline
\delta_{1} {  \mbf \mu} &  \ldots & \delta_{R+N_c} {  \mbf \mu}  \\
\end{array}
\right)
= -   \left(
\begin{array}{c| c}
{\bf E}_{R}&{\bf  0}\\\hline
{\bf 0}& {\bf E'}_{N_c}
\end{array}
\right),\label{Adx=-E}
\end{align}
where the vertical and horizontal lines indicate the structure of block matrices.
The quantities above are defined as follows: First, ${\bf E}_{R}$,   ${\bf E'}_{N_c}$ are  $R \times R$ and $N_c \times N_c$ diagonal matrices respectively, where the $\j$-th  component of ${\bf E}_{R}$ is given by $\frac{\partial  r_\j}{\partial k_\j} \delta k_\j$, and the $\a$-th component of ${\bf E}_{N_c}$ is given by  $\delta l_\a$. For a fixed $\J$, $\delta_\J \mbf x$ and $\delta_\J  \mbf \mu$ in \eqref{Adx=-E} are column vectors,  
\bal
\delta_\J \mbf x = (\delta_\J x_1,\ldots, \delta_\J x_M)^t, \nonumber\\
\delta_\J \mbf \mu = (\delta_\J \mu^1,\ldots, \delta_\J \mu^N)^t,
\eal
where $\delta_\J \mu^{\alpha}$ is the change of the coefficients in Eq. \eqref{kernel} under perturbation  $K_\J \rightarrow K_\J + \delta K_\J$. Finally, 
the matrix ${\bf A}$ is defined as 
\begin{align}
{\bf A}\equiv \begin{array}{c}
\\
\ _R\Bigg{\updownarrow} \\
\\
\ _{N_c}\Bigg{\updownarrow}\\
\\
\end{array}
\underset{\underset{M}{\xlongleftrightarrow[]{\hspace{4em}}} \ \ \ \ \ \ \underset{N}{\xlongleftrightarrow[]{\hspace{5em}}}}{\left(
\begin{array}{cccc|c }
&&&&  \\
&\mbox{\smash{ $r_{jm}$}}&&&{ - {\mbf c}^{\,1}\ {\ldots}\ - {\mbf c}^{\,N}}\\
&&&&\\\hline
&{-({\mbf d}^{\,1})^T}& & &\\ 
&\vdots&&&\mbox{\smash{\ $\bf 0$}} \\
&{-({\mbf d}^{\,N_{c}})^T}& & &
\end{array}
\right)},\label{Amat}
\end{align}
Note that the matrix $\bf A$ is proved to be square. 

By assuming that the matrix $\bf A$ is regular, Eq. \eqref{Adx=-E} uniquely determines the  sensitivity of chemicals, $\delta_\J x_m$,  and of the flux coefficients, $\delta_\J  \mu^\alpha$, as
 \bal
\left(
\begin{array}{cccc}
\delta_1 {\mbf  x} &  \ldots  &\delta_{R+N_c} {\mbf  x}  \\\hline
\delta_{1} { \mbf \mu} &  \ldots & \delta_{R+N_c} {  \mbf\mu}  \\
\end{array}
\right)
=   -{\bf A}^{-1} \left(
\begin{array}{c| c}
{\bf E}_{R}&{\bf  0}\\\hline
{\bf 0}& {\bf E'}_{N_c}
\end{array}
\right).\label{dx=A-1E}
\eal
By using  Eq. \eqref{kernel} and noting $\mbf c^\alpha$ is constant, $\delta_\J  \mu^\alpha$ determines the flux responses  as
\bal
\delta_\J r_j = \sum_{\alpha=1}^N \delta_\J\mu^\alpha c_j^\alpha,  
\eal
or, in matrix notation, 
\begin{align}
&\left(
\begin{array}{c c c}
\delta_{1} {\mbf r}&\ldots &\delta_{R+N_c} {\mbf r} 
\end{array}
\right)=\nonumber \\
&\left(
\begin{array}{c c c}
{\mbf c}^{\,1} &\ldots & { {\mbf c}^{\,N}}
\end{array}
\right)
\left(
\begin{array}{c c c }
\delta_1 {\mbf \mu}&\ldots &\delta_{R+N_c}{\mbf  \mu}
\end{array}
\right). \label{fluxresponse}
\end{align}
Note that  $\delta_\J {\mbf r}$ and ${\mbf c}^\alpha$ are $R$-dimensional column vectors, and $\delta_\J {\mbf \mu}$ are $N$-dimensional column vectors. 


Comments are in order. 
Firstly, practically, we are often interested in qualitative responses, (increased/decreased/invariant).
For such discussions,   assuming ``overexpressions" $\delta  K_J>0$, we can replace ${\bf E}_{R}$ and ${\bf E}'_{N_c}$ by identity matrices. 
Therefore, we call ${\bf S} \equiv -{\bf A}^{-1}$  {\it sensitivity matrix}. 
 Secondly, as a slight generalization,  we  can include nontrivial regulations such as allosteric effects by relaxing  Eq. \eqref{noreg} as
\begin{align}
\begin{cases}
\frac{\partial  {\color{black}r}_i}{\partial x_m}     \neq 0 \  \    {\rm  if}\  x_m \ {\rm influences \ reaction} \ i,  \\
  \frac{\partial {\color{black}r}_i}{\partial x_m} =0\  \  {\rm otherwise.}\ 
  \end{cases} 
  {\tag{2'}}
\label{reg}
\end{align}
Such regulations  add  additional nonzero $r_{im}$ in the $\bf A$-matrix, but  the response is still determined through Eq. \eqref{dx=A-1E}. Finally, although we are using the terminology ``perturbation",  the change of  parameters $\delta K_\J$ is not necessarily small for  qualitative discussion of the sensitivity as long as  the steady state persists for finite perturbations. 

\section{Example network}\label{sec:hierarchy}
We illustrate our method in  a simple example shown in FIG. \ref{fig:hierarchy-ex}, which has $N_c=1$. See  Appendix and  \cite{OM}  for examples with $N_c=0$.


 The stoichiometric matrix is given by
   \bal
   \nu = \left(
    \begin{array}{ccccc}
1&-1&-1&1& 0\\
0&1&-1&1& -1\\
0&0&1&-1& 0\\
0&0&1&-1& 0
    \end{array}
  \right).
\eal
 $\nu$ has a cokernel vector 
${\mbf d}_1 =(0,0,1,-1)$ because the difference $l_1\equiv {\mbf d}_1\cdot {\mbf x} = x_C - x_D$ is conserved in any reactions.
 $\bf A$ is given by
\bal
 {\bf A} = \left(
    \begin{array}{cccc|cc}
0 &0 & 0 &0 & 1 &  0\\
 r_{2A} &0 & 0 & 0&  1&0  \\
 r_{3A}  &  r_{3B} & 0&0 & 0 & 1 \\
 0   & 0& r_{4C} & r_{4D}& 0 & 1 \\
   0  & r_{5B}& 0& 0& 1 &  0\\ \hline
    0  & 0& 1&-1 &  0& 0  
    \end{array}
  \right).
\eal
From \eqref{Adx=-E}, the sensitivity is then calculated as
\bal
\delta_\J x_m &=\left(
\begin{array}{ccccc|c}
 \frac{1}{r_{2 A}} & \frac{-1}{r_{2 A}} & 0 & 0 & 0 &0\\
 \frac{1}{r_{5 B}} & 0 & 0 & 0 & -\frac{1}{r_{5 B}} &0\\
 \frac{R_2}{R_1 r_{2 A} r_{5 B}} & \frac{-r_{3 A}}{R_1 r_{2 A}} & \frac{1}{R_1} & \frac{-1}{R_1} & \frac{-r_{3 B}}{R_1 r_{5 B}} & \frac{r_{4D}}{R_1}\\
\frac{R_2}{R_1 r_{2 A} r_{5 B}} & \frac{-r_{3 A}}{R_1 r_{2 A}} & \frac{1}{R_1} & \frac{-1}{R_1} & \frac{-r_{3 B}}{R_1 r_{5 B}}&  \frac{-r_{4C}}{R_1}\\
\end{array}
\right)_{m\J},\nonumber\\
\delta_\J r_j &=
\left(
\begin{array}{ccccc|c}
 1 & 0 & 0 & 0 & 0&0 \\
 1 & 0 & 0 & 0 & 0&0  \\
 \frac{R_2}{r_{2 A} r_{5 B}} & -\frac{r_{3 A}}{r_{2 A}} & 1 & 0 & -\frac{r_{3 B}}{r_{5 B}} &0 \\
 \frac{R_2}{r_{2 A} r_{5 B}} & -\frac{r_{3 A}}{r_{2 A}} & 1 & 0 & -\frac{r_{3 B}}{r_{5 B}}&0  \\
 1 & 0 & 0 & 0 & 0&0  \\
\end{array}
\right)_{j\J}
\eal
where $R_1\equiv r_{4C} +r_{4D}$ and $R_2 \equiv r_{2A}r_{3B} + r_{3A}r_{5B}$. The $J^*$-th column is the response to perturbations of $J^*$-th parameter and the vertical lines separate between perturbations of the rate parameters and the conserved quantity. 
For example, from the fifth column, we can see that the increase of the initial  value of $l_1$ changes only $C$ and $D$, but does not affect  either $A $, $B$ or all fluxes. 

\begin{figure}[htbp]
    \includegraphics[width=8cm,bb=-40 0 300 150]{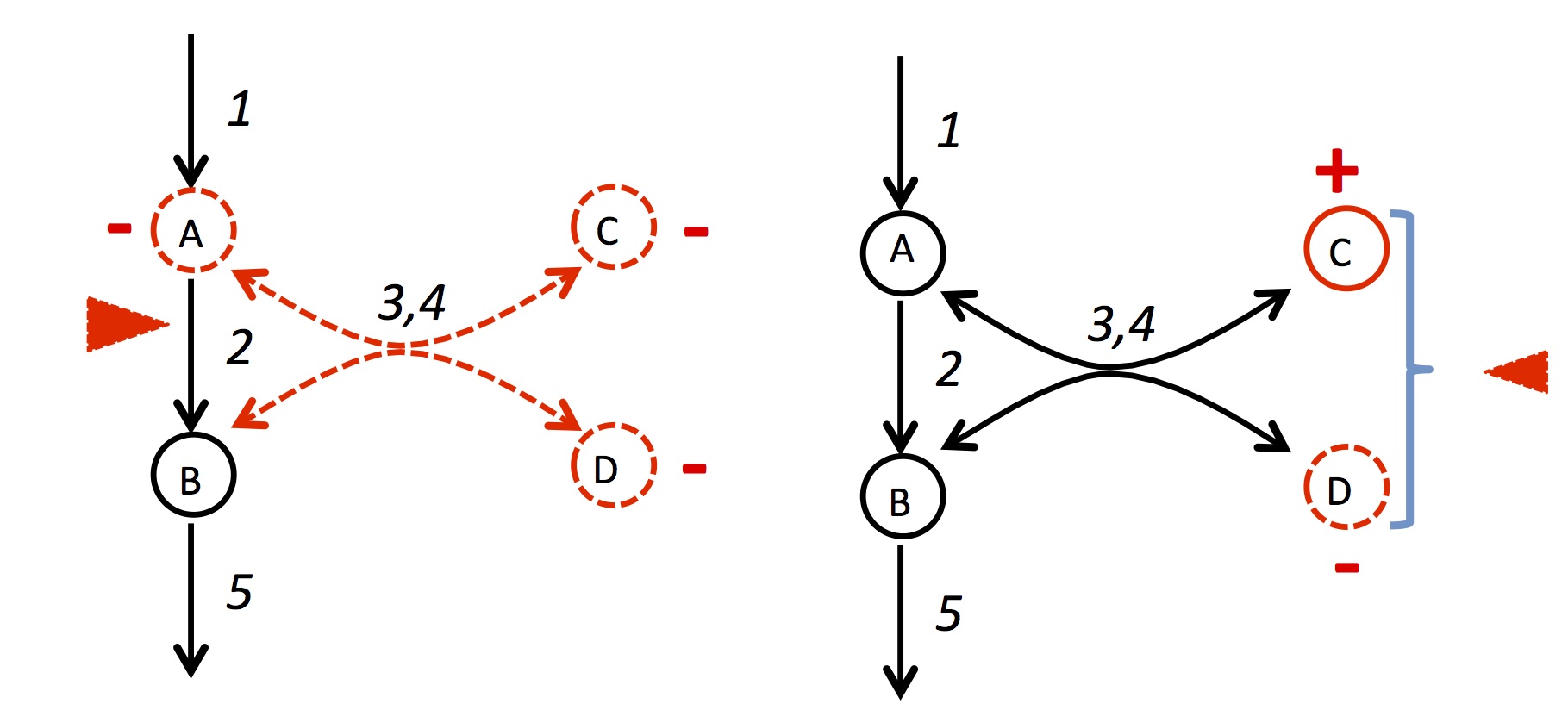}
  \caption{A network with  $N_c=1$. $1:(input) \rightarrow A,\ 2: A \rightarrow B, \ 3: A+B \rightarrow C+D, \ 4:  C+D \rightarrow A+B,\ 5: B\rightarrow (output)$. {\color{black} The responses of concentrations and fluxes are shown under the perturbations indicated by the red triangles.  The left-hand side figure shows that  increasing $k_2$ makes $x_A,\ x_B,\ x_C$ and $r_3,\ r_4$  decreased. The right-hand side figure shows that increasing  the initial value of $x_C - x_D$ makes $x_C$ increased and that of $x_D$ decreased. The other concentrations and all fluxes are not changed under the perturbation. }}
   \label{fig:hierarchy-ex}
\end{figure}

\section{Law of localization}\label{sec:localization}  
 As proved in \cite{OM}, for networks without conserved quantities, i.e. $N_c =0$, patterns of nonzero responses can be understood from network topology by using  a theorem called the law of localization. The theorem is also useful for elucidating relevant pathways by combining with experimental measurements.  
In this section, we generalize the theorem into networks  with conserved quantities, i.e. $N_c >0$.

First, we review the theorem for networks with $N_c =0$. For a given network, we consider a pair $\Gamma=({\mathfrak m}, {\mathfrak r})$  of a chemical subset ${\mathfrak m}$ and a reaction subset ${\mathfrak r}$ satisfying the condition that ${\mathfrak r}$ includes all reactions influenced by metabolites in ${\mathfrak m}$ [see the condition Eq. \eqref{reg}]. 
We  call  $\Gamma=({\mathfrak m}, {\mathfrak r})$ satisfying this condition {\it output-complete}. 
For a subnetwork $\Gamma=({\mathfrak m}, {\mathfrak r})$, we define an index,
 \bal
\lambda(\Gamma)\equiv- |{\mathfrak m}|+ |{\mathfrak r} |- N({\mathfrak r}). \label{LoL}
\eal
 Here, $|{\mathfrak m}|$  is  the number of elements in ${\mathfrak m}$, $|{\mathfrak r}|$  the number of elements in ${\mathfrak r}$, and  $N({\mathfrak r})$
the number  of independent {\it stoichiometric cycles} in ${\mathfrak r}$. By a stoichiometric cycle in ${\mathfrak r}$ we mean any flux vector $\mbf c$ which satisfies the flux balance $\nu \, {\mbf c}=0$ and nonzero components only within $\mathfrak r$. We call an output-complete subnetwork $\Gamma$   with  $\lambda(\Gamma)=0$ as a {\it buffering structure}. 



The  law of localization states that {\it   the  chemical concentrations and reaction fluxes outside of a buffering structure $\Gamma$ does not change under any  rate parameter perturbations in $\mathfrak r$.}  In other words, all effects of perturbations of  $k_\j$ in $\mathfrak r$ are  indeed localized within $\Gamma$. 
See Appendix for an illustration of the theorem for an example network with $N_c =0$.

Now, we state the theorem in  the case of $N_c \geq 0$ (see Appendix for the proof).   
We replace the definition \eqref{LoL} of the index $\lambda$ by 
\bal
\lambda(\Gamma)\equiv - |{\mathfrak m}|+ |{\mathfrak r} |- N({\mathfrak r}) +N_c({\mathfrak m}).\label{generalLoL}
\eal
The additional contribution, $N_c({\mathfrak m})$, is  the number of 
independent conserved quantities including at least one element in ${\mathfrak m}$. Note that the independencies of cycles and conserved quantities are defined in the vector spaces associated with  $\mathfrak r$ and ${\mathfrak m}$ respectively (see  Appendix B for a precise formal definition). 


The generalized law of localization then states that {\it  the  chemical concentrations and reaction fluxes outside of a buffering structure $\Gamma$ do not change under  either perturbations of   rate parameters or of conserved quantities in $\Gamma$}. 

We illustrate the generalized { law of localization} in the previous example network, shown in  FIG. \ref{fig:hierarchy-ex}, in which there is a conserved quantity $l_1 \equiv x_C-x_D$.  $N_c({\mathfrak m})=1$ if we choose ${\mathfrak m}$ including $C$ or $D$, such as ${\mathfrak m}=\{C\},\{D\}$, $\{C,D\}$, or  $\{A,C,D\}$. For the output-complete subgraph $\Gamma_1 = (\{C,D\},\{4\})$, the  index is  $\lambda(\Gamma_1)=-2+1-0+1=0$. For $\Gamma_2 = (\{C,D\},\{3,4\})$, which has one cycle consisting of reactions $3,\ 4$,  the index again vanishes; $\lambda(\Gamma_2)=-2+2-1+1=0$. Therefore, these subnetworks are buffering structures. This is consistent for the result that the perturbation of  the initial value of $x_C-x_D$ neither influences the concentrations of $A$ nor $B$. $\Gamma_3=(\{A,C,D\},\{2,3,4\})$ is another buffering structure with $\lambda(\Gamma_3)=-3+3-1+1$, which explains why $x_B$ is insensitive to the perturbations of $k_2,\ k_3,\ k_4$, and $l_1$. 

\section{Applications to biological networks}
We apply the structural sensitivity analysis and the law of localization to two signal transduction pathways which have  $N_c >0$.

\subsection{Signal transduction 1: MAPK}
\begin{figure}[htbp]
  \centering 
  \includegraphics[width=7cm,bb=0 0 300 220]{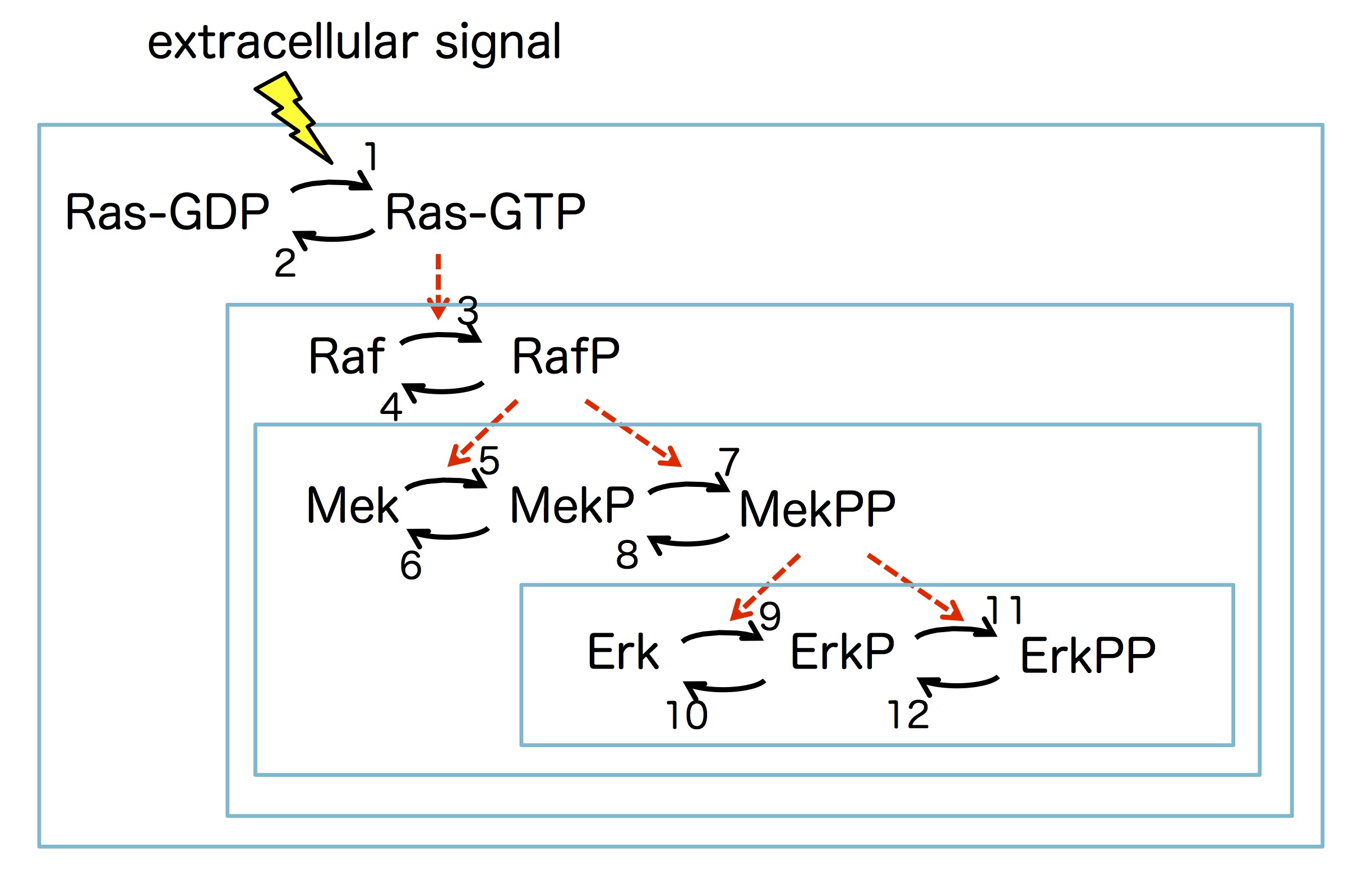}
   \caption{Signal transduction network of MAPK. The four boxes are four buffering structures. The solid lines represent state transitions of phosphorylations, and the dashed lines are active regulations.}   
   \label{fig:MAPK}
\end{figure}

We first consider the signal transduction network shown in FIG. \ref{fig:MAPK}. 
The stoichiometric matrix $\nu$ of this system has a four-dimensional cokernel space corresponding to the total amounts of Ras, Raf, Mek, and Erk. Phosphorylated chemicals in the upper layer positively regulate   phosphorylations in the lower one.
Mathematically, this means that, for example, the arguments of the flux function $r_7$ is not the form $r_7(k_7, {\rm MekP})$, but $r_7(k_7, {\rm MekP, RafP})$, which 
additionally adds the component $r_{7,\rm RafP}$ in the $\bf A$ matrix (see Eq. \eqref{reg}).

To construct  the matrix  $\bf A$, we order the chemicals as 
\bal
{\rm \{RasD,RasT,Raf,RafP,Mek,MekP},\nonumber\\{\rm MekPP,Erk,ErkP,ErkPP \}},
\eal
where RasD/RasT denotes the bound state of Ras and ADP/ATP, and choose the following basis for cokernel vectors;
\bal
{\mbf d}_{1}=(0,0,0,0,0,0,0,1,1,1)^T\nonumber\\
{\mbf d}_{2}=(0,0,0,0,1,1,1,0,0,0)^T\nonumber\\
{\mbf d}_{3}=(0,0,1,1,0,0,0,0,0,0)^T\nonumber\\
{\mbf d}_{4}=(1,1,0,0,0,0,0,0,0,0)^T\nonumber\\
\eal
\begin{widetext}
The  matrix  $\bf A$ is given by
{\scriptsize

\bal
\A=\left(
\begin{array}{cccccccccccc|cccc}
 r_{1, {\rm RasD}} & 0 & 0 & 0 & 0 & 0 & 0 & 0 & 0 & 0 & 0 & 0 & 0 & 0 & 0 & -1 \\
 0 & r_{2, {\rm RasT}} & 0 & 0 & 0 & 0 & 0 & 0 & 0 & 0 & 0 & 0 & 0 & 0 & 0 & -1 \\
 0 & r_{3, {\rm RasT}} & \cellcolor{LLLG} r_{3, {\rm Raf}} &  \cellcolor{LLLG}0 & \cellcolor{LLLG} 0 & \cellcolor{LLLG} 0 & \cellcolor{LLLG} 0 & \cellcolor{LLLG} 0 &  \cellcolor{LLLG}0 & \cellcolor{LLLG} 0 &  \cellcolor{LLLG}0 &  \cellcolor{LLLG}0 &  \cellcolor{LLLG}0 &  \cellcolor{LLLG}0 &  \cellcolor{LLLG}-1 & 0 \\
 0 & 0 & \cellcolor{LLLG} 0 & \cellcolor{LLLG} r_{4, {\rm RafP}} &  \cellcolor{LLLG}0 & \cellcolor{LLLG} 0 &  \cellcolor{LLLG}0 & \cellcolor{LLLG} 0 & \cellcolor{LLLG} 0 & \cellcolor{LLLG} 0 & \cellcolor{LLLG} 0 & \cellcolor{LLLG} 0 & \cellcolor{LLLG} 0 & \cellcolor{LLLG} 0 & \cellcolor{LLLG} -1 & 0 \\
 0 & 0 &  \cellcolor{LLLG}0 & \cellcolor{LLLG} r_{5, {\rm RafP}} & \cellcolor{LLG} r_{5, {\rm Mek}} &  \cellcolor{LLG}0 & \cellcolor{LLG} 0 &  \cellcolor{LLG}0 & \cellcolor{LLG} 0 &  \cellcolor{LLG}0 & \cellcolor{LLG} 0 & \cellcolor{LLG} 0 & \cellcolor{LLG} 0 &  \cellcolor{LLG}-1 &  \cellcolor{LLLG}0 & 0 \\
 0 & 0 &  \cellcolor{LLLG}0 &  \cellcolor{LLLG}0 & \cellcolor{LLG} 0 &  \cellcolor{LLG}r_{6, {\rm MekP}} & \cellcolor{LLG} 0 &  \cellcolor{LLG}0 & \cellcolor{LLG} 0 & \cellcolor{LLG} 0 &  \cellcolor{LLG}0 &  \cellcolor{LLG}0 &  \cellcolor{LLG}0 &  \cellcolor{LLG}-1 &  \cellcolor{LLLG}0 & 0 \\
 0 & 0 &  \cellcolor{LLLG}0 & \cellcolor{LLLG} r_{7, {\rm RafP}} &  \cellcolor{LLG}0 &  \cellcolor{LLG}r_{7, {\rm MekP}} &  \cellcolor{LLG}0 &  \cellcolor{LLG}0 & \cellcolor{LLG} 0 &  \cellcolor{LLG}0 &  \cellcolor{LLG}0 & \cellcolor{LLG} 0 &  \cellcolor{LLG}-1 &  \cellcolor{LLG}0 & \cellcolor{LLLG} 0 & 0 \\
 0 & 0 &  \cellcolor{LLLG}0 & \cellcolor{LLLG} 0 & \cellcolor{LLG} 0 & \cellcolor{LLG} 0 &\cellcolor{LLG}  r_{8, {\rm MekPP}} & \cellcolor{LLG} 0 &  \cellcolor{LLG}0 &  \cellcolor{LLG}0 &  \cellcolor{LLG}0 & \cellcolor{LLG} 0 &  \cellcolor{LLG}-1 & \cellcolor{LLG} 0 & \cellcolor{LLLG} 0 & 0 \\
 0 & 0 &  \cellcolor{LLLG}0 &\cellcolor{LLLG}  0 &\cellcolor{LLG}  0 & \cellcolor{LLG}0  &\cellcolor{LLG} r_{9, {\rm MekPP}} &\cellcolor{LG} r_{9, {\rm Erk}} &\cellcolor{LG} 0 &\cellcolor{LG} 0 & \cellcolor{LG}0 & \cellcolor{LG} -1 & \cellcolor{LLG} 0 &  \cellcolor{LLG}0 &  \cellcolor{LLLG}0 & 0 \\
 0 & 0 & \cellcolor{LLLG} 0 &\cellcolor{LLLG}  0 &\cellcolor{LLG}  0 &\cellcolor{LLG}  0 &\cellcolor{LLG} 0 & \cellcolor{LG} 0 & \cellcolor{LG}  r_{10, {\rm ErkP}} & \cellcolor{LG}  0 & \cellcolor{LG}0 & \cellcolor{LG} -1 &\cellcolor{LLG}  0 & \cellcolor{LLG} 0 & \cellcolor{LLLG} 0 & 0 \\
 0 & 0 & \cellcolor{LLLG} 0 & \cellcolor{LLLG} 0 &\cellcolor{LLG}  0 & \cellcolor{LLG} 0 & \cellcolor{LLG}  r_{11, {\rm MekPP}} & \cellcolor{LG}0 &\cellcolor{LG} r_{11, {\rm ErkP}} &\cellcolor{LG} 0 & \cellcolor{LG}-1 & \cellcolor{LG} 0 & \cellcolor{LLG} 0 &  \cellcolor{LLG}0 &  \cellcolor{LLLG}0 & 0 \\
  0 & 0 & \cellcolor{LLLG} 0 & \cellcolor{LLLG} 0 &\cellcolor{LLG}  0 &\cellcolor{LLG}  0 & \cellcolor{LLG}0 & \cellcolor{LG}0 & \cellcolor{LG}0 &\cellcolor{LG} r_{12, {\rm ErkPP}} &\cellcolor{LG} -1 &\cellcolor{LG}  0 &\cellcolor{LLG}  0 & \cellcolor{LLG} 0 & \cellcolor{LLLG} 0 & 0 \\ \hline
 0 & 0 & \cellcolor{LLLG} 0 & \cellcolor{LLLG} 0 &\cellcolor{LLG}  0 &\cellcolor{LLG}  0 & \cellcolor{LLG}0 &\cellcolor{LG} -1 &\cellcolor{LG} -1 & \cellcolor{LG}-1 &\cellcolor{LG} 0 & \cellcolor{LG}  0 & \cellcolor{LLG} 0 &  \cellcolor{LLG}0 &  \cellcolor{LLLG}0 & 0 \\
 0 & 0 & \cellcolor{LLLG} 0 & \cellcolor{LLLG} 0 &\cellcolor{LLG}  -1 &\cellcolor{LLG}  -1 &\cellcolor{LLG}  -1 &\cellcolor{LLG}  0 & \cellcolor{LLG} 0 &\cellcolor{LLG}  0 \cellcolor{LLG} &\cellcolor{LLG}  0 &\cellcolor{LLG}  0 &\cellcolor{LLG}  0 &  \cellcolor{LLG}0 & \cellcolor{LLLG} 0 & 0 \\
 0 & 0 & \cellcolor{LLLG} -1 & \cellcolor{LLLG} -1 & \cellcolor{LLLG} 0 & \cellcolor{LLLG} 0 & \cellcolor{LLLG} 0 & \cellcolor{LLLG} 0 & \cellcolor{LLLG} 0 & \cellcolor{LLLG} 0 & \cellcolor{LLLG} 0 & \cellcolor{LLLG} 0 & \cellcolor{LLLG} 0 & \cellcolor{LLLG} 0 & \cellcolor{LLLG} 0 & 0 \\
 -1 & -1 & 0 & 0 & 0 & 0 & 0 & 0 & 0 & 0 & 0 & 0 & 0 & 0 & 0 & 0 \\
\end{array}
\right).
\eal
}
Here, the gradations of color in $\A$ show  block matrices corresponding to four buffering structures (see \eqref{bs1} and \eqref{bs234} below). 
The signs of the components of the sensitivity matrix ${\bf S} =-{\bf A}^{-1}$ are determined as 
{\scriptsize
\bal
\begin{array}{c|cccccccccccc|cccc}
 &   {       k}_1 &   {       k}_2 &   {       k}_3 &   {       k}_4 &   {       k}_5 &   {       k}_6 &   {       k}_7 &   {       k}_8 &   {       k}_9 &   {       k}_{10} &   {       k}_{11} &   {       k}_{12} &   {       l}_1 &   {       l}_2 &   {       l}_3 &         l_4 \\ \hline
   {\rm RasD} & - & + & 0 & 0 & 0 & 0 & 0 & 0 & 0 & 0 & 0 & 0 & 0 & 0 & 0 & + \\
   {\rm RasT} & + & - & 0 & 0 & 0 & 0 & 0 & 0 & 0 & 0 & 0 & 0 & 0 & 0 & 0 & + \\
   {\rm Raf} & - & + & - & + & 0 & 0 & 0 & 0 & 0 & 0 & 0 & 0 & 0 & 0 & + & - \\
   {\rm RafP} & + & - & + & - & 0 & 0 & 0 & 0 & 0 & 0 & 0 & 0 & 0 & 0 & + & + \\
   {\rm Mek} & - & + & - & + & - & + & - & + & 0 & 0 & 0 & 0 & 0 &+  &-  &-  \\
   {\rm MekP} & \pm  & \pm  & \pm  & \pm  & + & - & - & + & 0 & 0 & 0 & 0 & 0  & +  & \pm & \pm \\
   {\rm MekPP} & + & - & + & - & + & - & + & - & 0 & 0 & 0 & 0 & 0& + & + & +  \\
   {\rm Erk} & - & + & - & + & - & + & - & + & - & + & - & + &+ &  - & - & -  \\
   {\rm ErkP} & \pm  & \pm  & \pm  & \pm  & \pm  & \pm  & \pm  & \pm  & + & - & - & + &+ &  \pm  & \pm  & \pm   \\
   {\rm ErkPP} & + & - & + & - & + & - & + & - & + & - & + & - & + & + & + & + \\
   {} &   &   &   &   &   &   &   &   &   &   &   &   &   &   &   &   \\
   {       r}_1 & + & + & 0 & 0 & 0 & 0 & 0 & 0 & 0 & 0 & 0 & 0 &  0 & 0 & 0 & + \\
   {       r}_2 & + & + & 0 & 0 & 0 & 0 & 0 & 0 & 0 & 0 & 0 & 0 &  0 & 0 & 0 & + \\
   {       r}_3 & + & - & + & + & 0 & 0 & 0 & 0 & 0 & 0 & 0 & 0 & 0 & 0 & + &+ \\
   {       r}_4 & + & - & + & + & 0 & 0 & 0 & 0 & 0 & 0 & 0 & 0 & 0 & 0 & + & + \\
   {       r}_5 & \pm  & \pm  & \pm  & \pm  & + & + & - & + & 0 & 0 & 0 & 0 & 0 &  + &  \pm  & \pm  \\
   {       r}_6 & \pm  & \pm  & \pm  & \pm  & + & + & - & + & 0 & 0 & 0 & 0 & 0 & + &  \pm  & \pm  \\
   {       r}_7 & + & - & + & - & + & - & + & + & 0 & 0 & 0 & 0 &0 &  + & + & + \\
   {       r}_8 & + & - & + & - & + & - & + & + & 0 & 0 & 0 & 0 & 0 & + & + & + \\
   {       r}_9 & \pm  & \pm  & \pm  & \pm  & \pm  & \pm  & \pm  & \pm  & + & + & - & + & +  & \pm  & \pm  & \pm  \\
   {       r}_{10} & \pm  & \pm  & \pm  & \pm  & \pm  & \pm  & \pm  & \pm  & + & + & - & +  & + & \pm  & \pm  & \pm  \\
   {       r}_{11} & + & - & + & - & + & - & + & - & + & - & + & + & + & + & + & + \\
   {       r}_{12} & + & - & + & - & + & - & + & - & + & - & + & + & + & + & + & + \\
\end{array},
\label{mapkresponse}
\eal
}
\end{widetext}
where  $+$, $-$ represent qualitative responses under perturbations associated with column indices, $k_1,\ldots, l_4$. The symbol $\pm$ means that the sign depends on quantitative values of $r_{im}$. 

The zero entries in \eqref{mapkresponse}  can be easily understood from the law of localization. 
The four square boxes in FIG. \ref{fig:MAPK}  indicate four buffering structures, forming a nested structure.
The smallest buffering structure is 
\bal
\Gamma_1=(\{\rm Erk,ErkP,ErkPP\},\{9,10,11,12\}).\label{bs1}
\eal
This subnetwork has one conserved quantity, $l_4=x_{\rm Erk}+x_{\rm ErkP}+x_{\rm ErkPP}$, and so $N_c(\mathfrak m)=1$. Therefore $\lambda(\Gamma_1)=-3+4-2+1=0$.  The law of localization then states that the perturbations of $k_8,k_9,k_{10},k_{11}$ and 
$l_1$ does not change the other part of the system, which explains the zeros appearing in the columns associated with  $k_{9,10,11,12}$ and $l_1$ in Eq. \eqref{mapkresponse}.

The remaining  buffering structures are
\bal
\Gamma_2&=\Gamma_1 \cup (\{  {\rm Mek,MekP,MekPP} \},\{5,6,7,8\})\nonumber\\
\Gamma_3&=\Gamma_2 \cup (\{  {\rm Raf,RafP} \},\{3,4\})\nonumber\\
\Gamma_4 &= \Gamma_3\cup(\{{\rm RasD,RasT}\},\{1,2\}). \label{bs234}
\eal
These buffering structures explain the zero entries, and in particular, the nest of them explains the stair-like nonzero pattern in Eq. \eqref{mapkresponse}.

The nest of buffering structures implies that perturbations to an upper layer influence the lower layers of the signal transduction pathway.


\subsection{Signal transduction 2: MAPK}
\begin{figure}[htbp]
  \includegraphics[width=6.cm,bb=0 0 500 400]{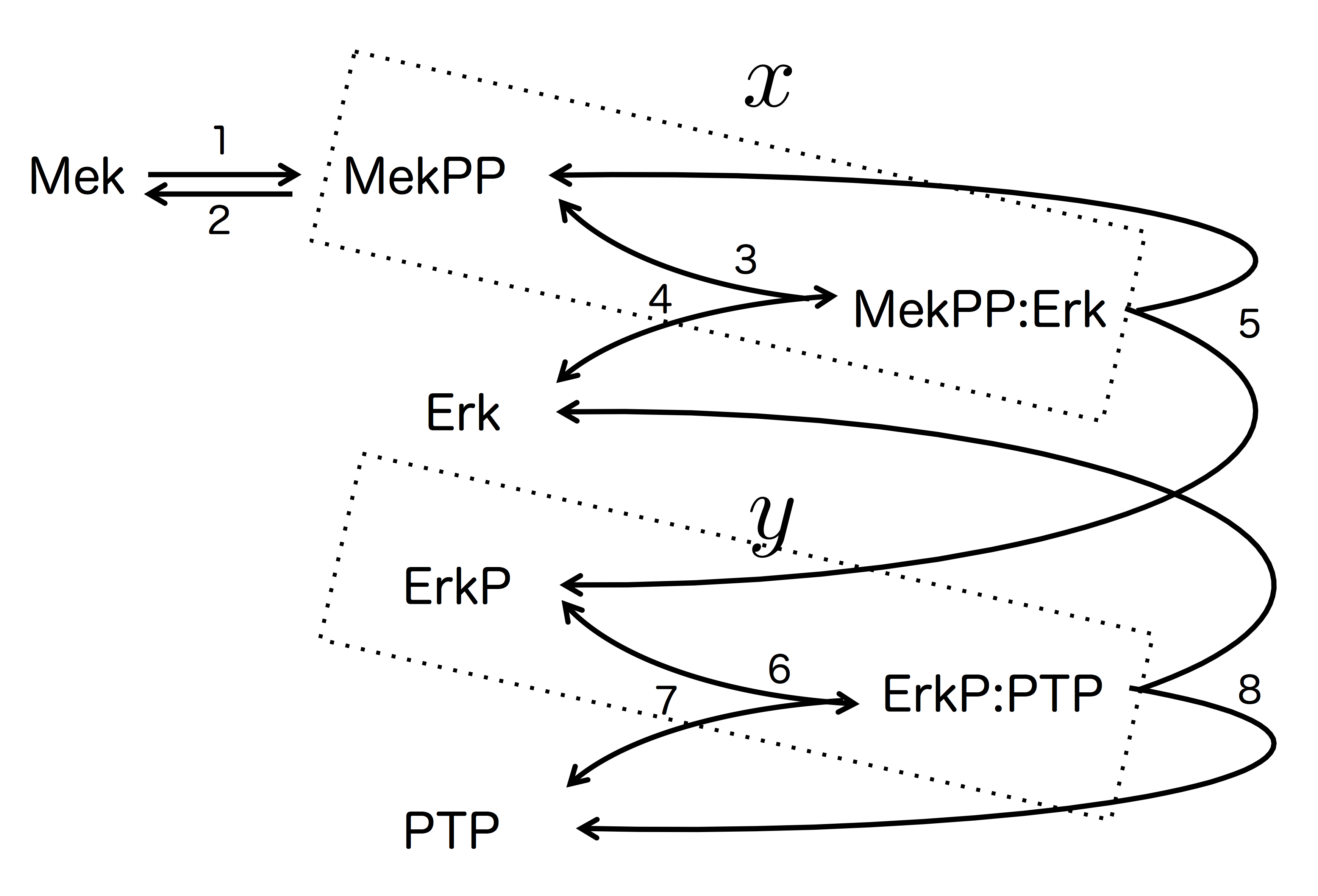}
   \caption{Signal transduction of MAPK/ERK pathway \cite{sontag}. 1: Mek $\rightarrow$ MekPP, 
   2: MekPP$\rightarrow$ Mek, 3: MekPP + Erk $\rightarrow$MekPP:Erk, 4: MekPP:Erk$\rightarrow$MekPP+Erk, 5: MekPP$\rightarrow$MekPP + ErkP, 6: ErkP + PTP $\rightarrow$ ErkP:PTP, {\color{black}7: ErkP:PTP $\rightarrow$ ErkP + PTP} 8: ErkP:PTP$\rightarrow$E{\color{black}r}k+PTP.   $x$ and $y$ denote $x_{\rm MekPP} + x_{\rm MekPP:Erk}$ and $x_{\rm ErkP}+x_{\rm ErkP:PTP}$ respectively.}   
   \label{fig:MAPKcomplex}
\end{figure}

{\color{black}We next study the signal transduction network shown in FIG. \ref{fig:MAPKcomplex}. This network was studied in \cite{sontag}, where the regulation between the bottom two layers in FIG. \ref{fig:MAPK} was modeled in detail as bound state formation in FIG. 4. They studied the sensitivity by using a clever manipulation of equations under the assumption of the mass-action kinetics. Here, by using structural sensitivity analysis, we derive the same result without assuming specific kinetics, which illustrates generality and usefulness of our method. We remark that, although this example is superficially similar to the one in FIG. \ref{fig:MAPK}, surprisingly, the result turns out to be more complex than the previous example.}

This system has three conserved quantities associated with the total amounts of Erk, PTP, and Mek. We order the chemicals as 
\bal
\{\rm Mek,MekPP,Erk,MekPP:Erk,\nonumber\\ \rm  ErkP,PTP,ErkP:PTP \},
\eal
and choose the basis of the cokernel space as 
\bal
{\mbf d}_1&=(0,0,1,1,1,0,1)^T\nonumber\\
{\mbf d}_2&=(0,0,0,0,0,1,1)^T\nonumber\\
{\mbf d}_3&=(1,1,0,1,0,0,0)^T.
\eal\begin{widetext}
The $\bf A$ matrix is given by
{\scriptsize
\bal
{\bf A}=\left(
\begin{array}{ccccccc|cccc}
 r_{1, {\rm Mek}} & 0 & 0 & 0 & 0 & 0 & 0 & 0 & 0 & 0 & -1 \\
 0 & r_{2, {\rm MekPP}} & 0 & 0 & 0 & 0 & 0 & 0 & 0 & 0 & -1 \\
 0 & r_{3, {\rm MekPP}} & r_{3, \rm {Erk}} & 0 & 0 & 0 & 0 & -1 & 0 & -1 & 0 \\
 0 & 0 & 0 & r_{4, {\rm MekPP:Erk}} & 0 & 0 & 0 & 0 & 0 & -1 & 0 \\
 0 & 0 & 0 & r_{5, {\rm MekPP:Erk}} & 0 & 0 & 0 & -1 & 0 & 0 & 0 \\
 0 & 0 & 0 & 0 & r_{6, {\rm Erkp}} & r_{6, {\rm PTP}} & 0 & -1 & -1 & 0 & 0 \\
 0 & 0 & 0 & 0 & 0 & 0 & r_{7, {\rm Erkp:PTP}} & 0 & -1 & 0 & 0 \\
 0 & 0 & 0 & 0 & 0 & 0 & r_{8, {\rm Erkp:PTP}} & -1 & 0 & 0 & 0 \\\hline
 0 & 0 & -1 & -1 & -1 & 0 & -1 & 0 & 0 & 0 & 0 \\
 0 & 0 & 0 & 0 & 0 & -1 & -1 & 0 & 0 & 0 & 0 \\
 -1 & -1 & 0 & -1 & 0 & 0 & 0 & 0 & 0 & 0 & 0 \\
\end{array}
\right)
\eal
}
\end{widetext}
The signs of responses are
{\scriptsize
\bal
\begin{array}{c|cccccccc|ccc}
 \rm{} & k_1 & k_2 & k_3 & k_4 & k_5 & k_6 & k_7 & k_8 & l_1 & l_2 & l_3 \\ \hline
 \rm{Mek} & - & + & - & + & + & - & + & - & - & - & + \\
 \rm{MekPP} & + & - & - & + & + & - & + & - & - & - & + \\
 \rm{Erk} & - & + & - & + & \pm & + & - & + & + & + & - \\
 \rm{MekPP:Erk} & + & - & + & - & - & + & - & + & + & + & + \\
 \rm{Erkp} & + & - & + & - & + & - & + & \pm & + & - & + \\
 \rm{PTP} & - & + & - & + & - & - & + & + & - & + & - \\
 \rm{Erkp:PTP} & + & - & + & - & + & + & - & - & + & + & + \\
 \rm{} &   &   &   &   &   &   &   &   &   &   &   \\
r_1 & + & + & - & + & + & - & + & - & - & - & + \\
r_2 & + & + & - & + & + & - & + & - & - & - & + \\
r_3 & + & - & + & + & \pm & + & - & + & + & + & + \\
r_4 & + & - & + & + & - & + & - & + & + & + & + \\
r_5 & + & - & + & - & + & + & - & + & + & + & + \\
r_6 & + & - & + & - & + & + & + & \pm & + & + & + \\
 r_7 & + & - & + & - & + & + & + & - & + & + & + \\
 r_8 & + & - & + & - & + & + & - & + & + & + & + \\
\end{array}
 \label{sontagresponse}\eal
}
Note that this network has only one trivial buffering structure, i.e. the whole network. Accordingly, there are no vanishing entries in \eqref{sontagresponse}.

Following \cite{sontag}, we focuse on total active Mek and  Erk concentrations, 
\bal
x&\equiv x_{\rm MekPP} + x_{\rm MekPP:Erk}, \nonumber\\ y &\equiv x_{\rm ErkP}+x_{\rm ErkP:PTP},
\eal
 and examine  their responses to the perturbations of $l_1,l_2$, which correspond to the total amounts of Erk and PTP with {\it any form}.
In our method,  these responses can be obtained by summing the associated rows in Eq. \eqref{sontagresponse}, which leads to 
\bal
\delta_{l_{1}}x&\propto r_{3,{\rm Erk}} r_{6,{\rm ErkP}} r_{8,{\rm ErkP:PTP}} r_{2,{\rm MekPP}}\nonumber\\
\delta_{l_{1}}y&\propto r_{3,{\rm Erk}} r_{5,{\rm MekPP:Erk}} \left(r_{1,{\rm Mek}}+r_{2,{\rm MekPP}}\right) \nonumber\\&\left(r_{6,{\rm ErkP}}+r_{7,{\rm ErkP:PTP}}+r_{8,{\rm ErkP:PTP}}+r_{6,{\rm PTP}}\right)\nonumber\\
\delta_{l_{2}}x&\propto r_{3,{\rm Erk}} r_{8,{\rm ErkP:PTP}} r_{2,{\rm MekPP}} r_{6,{\rm PTP}}\nonumber\\
\delta_{l_{2}}y&\propto- r_{6,\rm PTP} r_{8,\rm ERkP:PTP}\biggl( r_{2,\rm MekPP}(r_{3,\rm Erk}\nonumber\\ & +r_{4,\rm MekPP:Erk} + r_{5,\rm MekPP:Erk})+ r_{1,\rm Mek}(r_{3,\rm Erk}\nonumber\\ &+r_{3,\rm MekPP}+r_{4,\rm MekPP:Erk} +r_{5,\rm MekPP:Erk}) \biggr).
\eal
Here we omit the common  positive proportional constant, $({\rm Det}\, {\bf A})^{-1}>0$. We thus  obtain the following qualitative responses,
\begin{subequations}
 \label{sontagres}
 \begin{align}
\delta_{l_{1}}x>0,\ \delta_{l_{1}}y>0, \label{sontag1}\\
\delta_{l_{2}}x>0,\ \delta_{l_{2}}y<0\label{sontag2},
 \end{align}
\end{subequations}
which agrees with the result obtained in  \cite{sontag}.  The authors of \cite{sontag} called the result  \eqref{sontagres}  ``paradoxical results": While Eq. \eqref{sontag1}  suggests that $x$ activates $y$,
Eq. \eqref{sontag2}  suggests that $x$ inhibits $y$. 

We  emphasize that while the argument of \cite{sontag} is based on the mass-action type kinetics, we obtained the same   conclusion valid for general (monotonically increasing) flux functions.
Also, our systematic approach determines all responses simultaneously. 
We summarize the responses of $x,\, y$ for all perturbations;
\bal
\begin{array}{c|ccccccccccc}
 & k_1 & k_2 & k_3 & k_4 & k_5 & k_6 & k_7 & k_8 & l_1 & l_2 & l_3 \\ \hline
x & + & - & + & - & - & + & - & + & + & + & + \\
 y & + & - & + & - & + & - & + & - & + & - & + \\\end{array}
 \label{sontagresponse2}\eal

\section{Characteristics for robustness based on network structures}

\begin{figure}[htbp]
  \includegraphics[width=6.cm,bb=0 0 205 225]{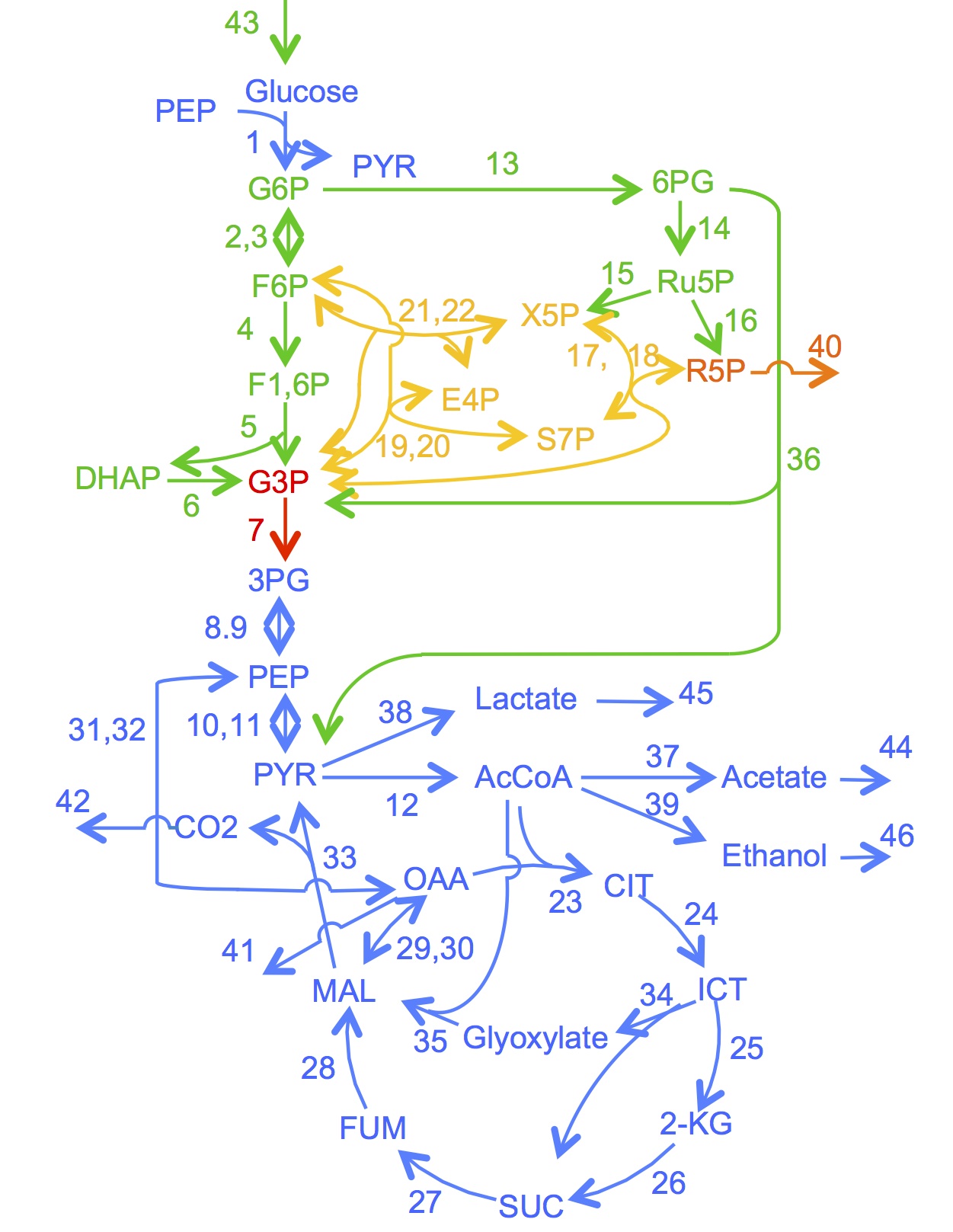}
   \caption{ The central metabolism network of E. coli, consisting of $M=28$ metabolites and $R=48$ reactions. (Adopted from \cite{Ishii}).  See Appendix for the list of reactions. 
  }
  \label{fig:zu1}
  \end{figure}
  \begin{figure}[htbp]
  \includegraphics[width=8.cm,bb=0 0 400 130]{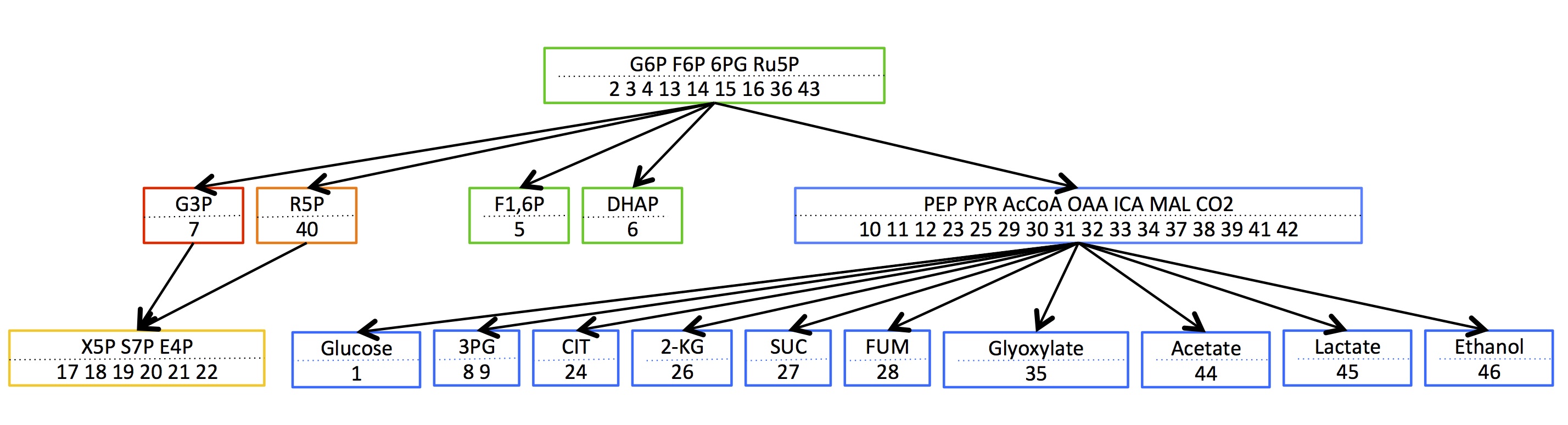}
   \caption{ 17 buffering structures of E. coli network.  Each box corresponds to each buffering structure. For any box, a union of metabolites and reactions in the box and boxes below it gives a buffering structure. For example,  $ \{\rm X5P, S7P, E4P, 17,18,19,20,21,22\} $ is a buffering structure, and a union of $\{ \rm G3P,7\}$ and $ \{\rm X5P, S7P, E4P, 17,18,19,20,21,22\} $ is another buffering structure.  }
   \label{fig:zu2}
\end{figure}

 In our previous study, we examined the carbon metabolism pathway of E. coli \cite{Mochizuki, OM} shown in FIG. \ref{fig:zu1}, which  is a major part of energy acquisition process. We found that the network has 17 buffering structures shown in   FIG. \ref{fig:zu2} (see also Appendix for the  list), and, in particular, that some of buffering structures coincide with subnetworks associated with biological functions such as the TCA cycle and the pentose phosphate pathway. These observations suggest that  biological networks are selected in evolution and include many buffering structures, which provide  robustness to enzymatic perturbations.   In this section,  in order to support  this expectation, we compare robustness property between the  E. coli network  and artificial random networks.

Firstly, we introduce network characteristics that quantify  robustness for  reaction systems. One natural definition is the number of buffering structures, $N_{BS}$, which is more precisely defined as {\it the number of buffering structures consisting of different sets of  chemicals}; {\color{black} we distinguish two buffering structures if they have at least one different metabolite. Note that we identify two buffering structures that have different sets of reactions even if they have  the same metabolite set}. 
 Another one is the fraction of metabolites  that  exhibit zero responses under a randomly chosen enzymatic perturbation ($k_\j \rightarrow k_\j + \delta k_\j$);
\bal
{\mathscr R} \equiv \frac{1}{M\times R} \sum^{R,M}_{\underset{ s.t.  \delta_\j x_m = 0}{\j=1,m=1 }} 1. \label{robustness}
\eal
By definition $0\leq {\mathscr R} \leq 1$.  
Networks with larger ${\mathscr R}$ and larger $N_{BS}$ are more robust. We emphasize that ${\mathscr R}$ and $N_{BS}$  are purely structural  characteristics determined from network topology alone. 
 

\begin{figure}[htbp]
  \centering
  \includegraphics[width=3.cm,bb=0 0 150 140]{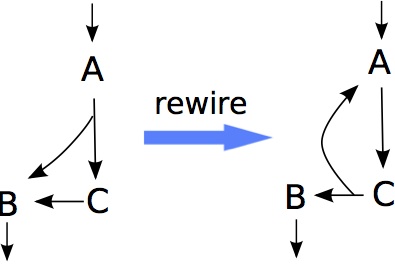}
   \caption{An illustration of rewiring procedure. Here the reaction $A\rightarrow B+C$ becomes  $C\rightarrow A + B$, and the reaction $C\rightarrow B$ becomes  $A\rightarrow C$. }  
   \label{fig:rewire-example}
\end{figure}

In order to generate random networks, we randomly choose    $p \times R=48p$ reactions of  the E. coli network and reconnect them to randomly chosen nodes of metabolites (see FIG. \ref{fig:rewire-example}) \cite{alon}. Here,  $p$ ($0 <p \leq1$) is a fraction of rewired reactions.
{\color{black}This is implemented by randomly choosing a column of the $M\times R$ stoichiometry matrix $\nu$, randomly reordering the $M$ components in the column, and repeat this procedure $48p$ times. } 
Rewired networks  are sometimes singular, i.e. ${\rm det}\,  \A = 0$, which cannot admit steady state, and sometimes unconnected, which is not suitable to compare with the E. coli network. We discard such networks. In this way, 
 we constructed an ensemble of  3000 regular and connected networks for each value of $p$. 


%

\begin{figure*}[!htb]
  \includegraphics[width=20cm,bb=0 0 500 140]{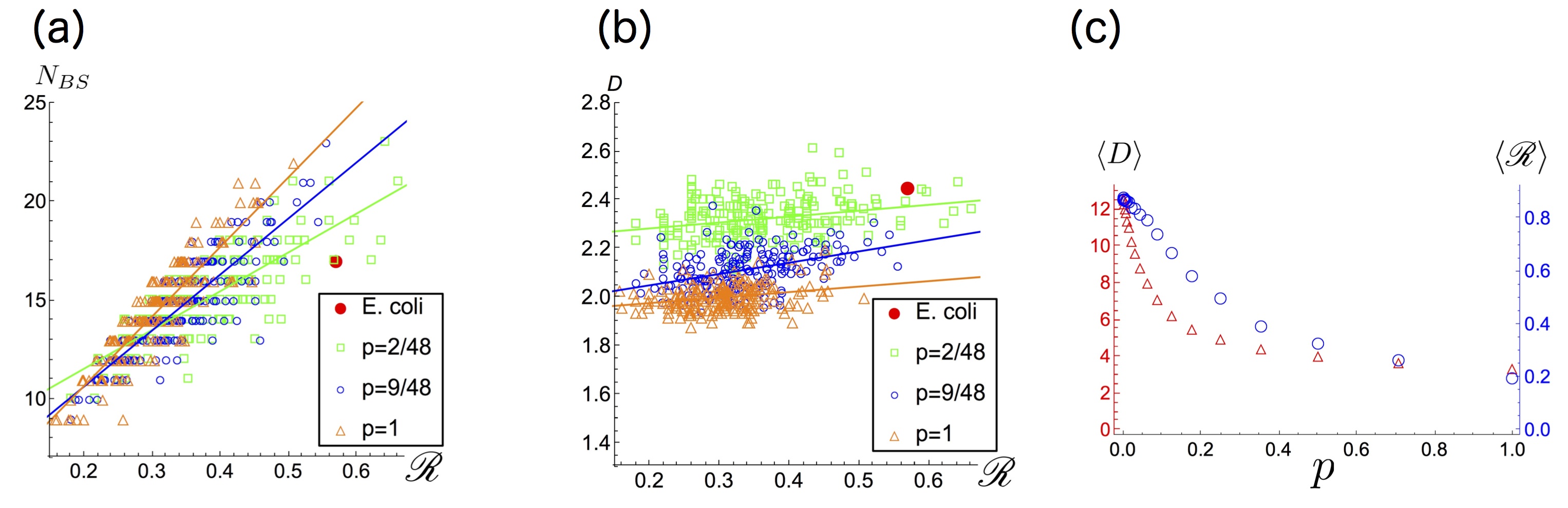}
  \caption{(a) The distribution of $({\mathscr R},N_{BS})$ of 200 sampled networks with  $p=2/48,\, 9/48$ and $1$. The filled circle represents the E. coli network. Pearson correlation coefficient (averaged over ensembles with $0<p \leq 1$) is $0.86$. (b)  The distribution of $({\mathscr R},D)$ of 200 sampled networks.  Pearson correlation coefficient is $0.32$. (c) The  averages of ${\mathscr R}$ and $D$ over  network ensemble with fixed $p$ ($0 <p \leq1$) are shown by circles and triangles respectively. The plots for $p=0$ correspond to the (unrewired) E. coli network. }
  \label{fig:random}
\end{figure*}

For these ensembles, we examined the robustness, ${\mathscr R}$ and $N_{BS}$, and the mean distance $D$, which is one of the most widely used network characteristics.  Here,  for a given network, the distance between two metabolites is defined as the smallest number of reactions that connects the two, and $D$ is the average value of the distances. {\color{black}More rigorously, $D$ here is defined for undirected networks where nodes represent metabolites and edges between two metabolites are drawn if the two metabolites are involved with the same reaction; for example,}  the distances between substrates and products of a reaction are defined as one.  

The results are as follows. 
Firstly, FIG. \ref{fig:random} (a) is the distributions of ${\mathscr R}$ and $N_{BS}$ for $p=2/48,\, 9/48,\, 1$. The filled red circle represents the E. coli network, ${\mathscr R}_{E.\, coli} \simeq 0.57$ and $N_{BS}\bigr|_{E.\, coli} \simeq 17$. We can see  a strong correlation between ${\mathscr R}$ and $N_{BS}$. Thus,  the robustness based on ${\mathscr R}$ and  $N_{BS}$ agree with each other statistically.
 
Secondly,  FIG. \ref{fig:random} (b) shows the distribution of ${\mathscr R}$ and  $D$ of a network. The blue circle represents the E. coli network, ${\mathscr R}_{E.\, coli} \simeq 0.57$ and $D_{E.\, coli} \simeq 2.45$. We can see that the distribution of ${\mathscr R}$ with fixed $D$ is  widely spread for any $p$. This means  that there are no significant correlations between ${\mathscr R}$ and $D$. 

Finally, FIG. \ref{fig:random} (c) is the ensemble average of ${\mathscr R}$ and $D$.   As we randomize the E.coli network by increasing $p$ from $0$ to $1$,   the robustness $ \langle {\mathscr R} \rangle$ and the mean distance $\langle D \rangle$ tend to decrease monotonically. Here, $\langle \cdot \rangle$ denotes the average over the network ensemble with fixed $p$.
We note that this tendency of  $ \langle {\mathscr R} \rangle$ and  $\langle D \rangle$
does {\it not} necessarily mean that there is a correlation between ${\mathscr R}$ and $D$. In fact, we did not find such a correlation in FIG. \ref{fig:random} (b).  
We also confirmed these behaviors in a model like the Watts-Strogatz model \cite{WS} (see Appendix). 
 
 One of the most remarkable results in  FIG. \ref{fig:random} (a) is the peak 
  of ${\mathscr R}$ at $p=0$, corresponding to the unrewired E. coli network. 
  This peak means that even a small fraction of rewiring (for example $p=2/48$) lowers the robustness ${\mathscr R}$ drastically (from ${\mathscr R}_{E.\, coli} \simeq 0.57$ to  $ \langle {\mathscr R} \rangle \simeq 0.37 $). {\color{black}The extraordinary robustness of the E. coli network suggests that the special topology, which is characterized by buffering structures, might be formed and selected under evolutionary pressures on the robustness. }
  

While the robustness $ \langle {\mathscr R} \rangle$ has a steep peak at $p=0$, $\langle D \rangle $ changes smoothly around $p=0$, as expected. In fact, other network quantities such as centrality, and  degree distributions also change smoothly around $p=0$. The uncorrelated distributions shown in FIG. \ref{fig:random} (b) and the sharp peak of ${\mathscr R}$ at $p=0$  imply that ${\mathscr R}$ and $N_{BS}$ are completely novel  characteristics for  robustness, which cannot be captured by any conventional graph theoretical quantities.

\section{Conclusions }
In this paper, we  generalized our previous formalism of structural sensitivity analysis and the law of localization into reaction systems with conserved quantities. Our generalized method can be applied into any biochemical systems, including signal transduction networks, metabolic systems, and protein synthesis, if the systems admit steady states. 

We   applied our method into  two signal transduction networks with conserved quantities. While  the authors in  \cite{sontag} studied the second network by assuming mass-action kinetics,  we obtained the same conclusion as theirs  {\it without assuming specific  kinetics} such as mass-action types and the Michaelis-Menten kinetics. This illustrates how powerful and general our method is. 

Our structural approach  is also practically useful in experimental biology.  In spite of the  progress in biosciences, it is difficult to experimentally determine  kinetics of biochemical reactions in living cells. 
Our method overcomes the difficulty because we determine qualitative sensitivity (increased/decreased/invariant) to perturbations only from network structures. By making use of this advantage,  we can  testify database information on networks systematically (see \cite{OM} for more detailed discussions).

Finally, we  investigated biological meanings of buffering structures by comparing E. coli network with  random networks. We  introduced two network characteristics  measuring robustness of networks;  the number of buffering structures $N_{BS}$ and  the fraction of zero responses to perturbations $\mathscr R$. Based on them, we  measured robustness of reactions systems. 
We  found that   even a partial rewiring deteriorates the robustness of the E. coli network drastically. E. coli network   has special features that cannot be captured by other indexes than ours and realize extraordinarily robust compared with random networks.  {\color{black}Our result suggests that the topology of the E. coli network might be selected under evolutionary pressures on robustness.}

The proposed quantities for robustness, $N_{BS}$ and $\mathscr R$, are completely novel network characteristics  because they are not correlated with conventional network characteristics, such as mean distance or degree distributions {\color{black}(see also Appendix)}.   We will study more analytical aspects  about the relation between robustness ($N_{BS}$ and $\mathscr R$) and network topology and hope to report on them in the near future.



This work was supported partly by the CREST, Japan Science and Technology Agency, and by iTHES Project/iTHEMS Program RIKEN, by Grant-in-Aid for Scientific Research on Innovative Area, “Logics of Plant Development,” Grant No. 25113001. We  express our sincere thanks to Testuo Hatsuda, Michio Hiroshima, Yoh Iwasa,  Masaki Matsumoto, Keiichi Nakayama, and Yasushi Sako for their helpful discussions and comments.



\appendix

\section{ Derivation of Equation \eqref{dx=A-1E}}\label{app:response}
 As in many  mathematical studies in metabolism like flux balance analysis 
 \cite{FBA1, FBA2, FBA3}, we are  focusing on  steady state, which is characterized by  
 \bal
0 = \sum_{j=1}^R \nu_{mj }r_j(k_j,\bar{x})\label{before}.
\eal
Here, $ \bar{x}_m$ represents the concentration of metabolite $m$ at  steady state, which generally depends on (a subset of) the parameters $\{ k_j \}$ and  initial condition specified by $\{ l_a\}$. We determine the sensitivity of each concentration and flux to the perturbations $k_\j \rightarrow k_\j +\delta k_\j$ and $l_a\rightarrow l_a+ \delta l_a$. 

First, we determine the responses to   perturbations of rate parameters.
We choose one reaction $j=j^*$ in the system and  perturb the  parameter as $k_{j^*}\rightarrow k_{j^*}+\delta k_{j^*}$.  
The system 
goes to a new steady state, characterized by
\bal
0 = \sum_{j=1}^R \nu_{mj }r_j(k_j+\delta_{j j^*}\delta k_{j^*},\bar{x}+\delta_{j^*} {x})\label{after}.
\eal
Here, $\delta_{j j^*}$ is the Kronecker delta symbol.  Below we determine the concentration sensitivity  $\delta_{j^*}  x_m $ and the flux sensitivity $\delta_{j^*} r_j \equiv  r_j(k_j+\delta_{j j^*}\delta k_{j^*},\bar{x}+\delta_{j^*} {x})  -r_j(k_j,\bar{x})$.

The above two equations imply that the flux change  also satisfies $\sum_j \nu_{mj} \delta_{j^*}r_j =0$. Therefore,  the vector $\delta_{j^*} {r} \in {\mathbb R}^R$ can be expanded in terms of a basis $\{ {\mbf c}^{\alpha}\}_{ \alpha=1,\ldots,N}$ of the kernel of $\nu$, where ${{\mbf c}^{\alpha}}\in {\mathbb R}^R $ and $N\equiv {\rm dim\, ker}\nu$;
 \bal
 \delta_{j^*} r_j   =\sum_{\alpha=1}^{N}\delta_{j^*}  \mu_{\alpha} \ {c}^{\alpha}_j.
 \label{dw1}
 \eal
 Here,  ${c}^{\alpha}_j$ is the $j$-th component of the kernel vector ${{\mbf c}^{\alpha}} $. Thus, the problem of determining the fluxes is equivalent to that of determining the coefficients  $\delta_{j^*} \mu_{\alpha}$ of the kernel vectors.

As we commented in the main text,  for the purpose of determining qualitative responses, we can assume that the perturbations  are small.  In the limit of infinitesimal perturbations $\delta k_\j$, Taylor expansion of   $ \delta_{j^*} r_j$ yields
\bal
\delta_{j^*} r_j  =\frac{\partial r_j}{\partial k_j}\biggr|_{x= \bar{x}} \delta k_{j^*}  \ \delta_{j,j^*}+ \sum_{m=1}^M r_{jm}\, \delta_{j^*}  x_m , \label{dw2}
 \eal
where we abbreviate  $r_{jm} \equiv \frac{\partial r_j}{\partial x_m}\bigr|_{x= {\bar x}}$.
Comparing Eqs. \eqref{dw1} and \eqref{dw2}, we obtain
 \bal
 \sum_{m=1}^M r_{jm}\,\delta_{j^*}  x_m
 -\sum_{\alpha=1}^{N} {c}^{\alpha}_j  \,   \delta_{j^*} \mu_{\alpha}=- 
\frac{\partial r_j}{\partial k_j}\biggr|_{x= {\bar x}} \delta k_{j^*}  \ \delta_{j,j^*}. \label{cond1}
 \eal

When  the cokernel vectors of $\nu$ exist, i.e.  $ {\rm dim\ coker}\, \nu \equiv N_c >0$  , Eq. \eqref{cond1} is not enough to determine the sensitivity, and we need additional constraints on the concentration changes. Let the constant vectors $\{ {\mbf d}^{\,a}  \}_{a=1,\ldots, N_c}$ be a basis of the cokernel space, where ${\mbf d}^a \in {\mathbb R}^R $, and $N_c$ is the dimension of the cokernel space. Then  the linear combinations 
\bal
l_a \equiv \sum_{m=1}^M ({\mbf d}_{a})_m x_m (t)\, \ \ (a=1,\ldots,  N_c ),
\eal
where $({\mbf d}_{a})_m$ denotes the $m$-th component of ${\mbf d}_a$, are conserved in the dynamics of Eq. \eqref{ode}. This implies that the perturbed steady state depends on $\{ l_a\}$.
In order to make the problem of the sensitivity well-defined, we assume  the perturbed system starts  with the same initial condition as the unperturbed system.  
Then the concentration changes need to satisfy
\bal
\sum_{m=1}^M ({\mbf d}_{a})_m\,  \delta_{j^*}  x_m=0\label{cond2}.
\eal
for all $a=1,\ldots, N_c$.

Eqs. \eqref{cond1} and \eqref{cond2} determine the response to the perturbation $\j$. In matrix notation, these can be written as
\begin{eqnarray}
{\bf A}\left(
\begin{array}{c}
\delta_{j^*} { \mbf x} \\ \hline
\delta_{j^*} {\mbf \mu}
\end{array}
\right)\begin{array}{c }
{\updownarrow}\ _M \\
{\updownarrow}\ _{N}\\
\end{array}
= - \left(
\begin{array}{c}
{ {\mbf e}_{j^*} }\\\hline
{\bf 0}
\end{array}
\right)\begin{array}{c }
{\updownarrow}\ _R \\
{\updownarrow}\ _{N_c}\\
\end{array}\label{appwocoker},
\end{eqnarray}
where the matrix $\bf A$ is defined in Eq. \eqref{Amat}, and  $ {\mbf e}_{j^*} \equiv (0,\ldots,\underset{{j^*}{\rm{\mathchar`-}th}}{\underbrace{\frac{\partial r_{j^*}}{\partial k_{j^*}}\biggr|_{x= \bar x} \delta k_{j^*}}} ,\ldots,0 )^T\in{\mathbb R}^R$. 

We note that the matrix $\bf A$ is square, namely the identity $M + N = R + N_c$ holds.  This follows from the well-known identity for the Fredholm index, ${\rm dim\, ker}\, \nu - {\rm dim \,coker}\, \nu = R - M$ for any $M\times R$ matrix $\nu: {\mathbb R}^R \rightarrow {\mathbb R}^M$.


Next, we discuss responses to perturbation of conserved quantities, or initial concentrations.
 We choose a particular conserved quantity, $a=a^*$, and consider the perturbation $l_{a^*} \rightarrow l_{a^*} + \delta l_{a^*}$.
Eqs. \eqref{dw1}, \eqref{dw2}, \eqref{cond1}, and \eqref{cond2} are replaced by
 \bal
 \delta_{a^*} r_j   =\sum_{\alpha=1}^{N}\delta_{a^*}  \mu_{\alpha} \ {c}^{\alpha}_j,
 \eal
 \bal
\delta_{a^*} r_j  = \sum_{m=1}^M r_{jm}\, \delta_{a^*}  x_m , \label{dw2'}
 \eal
 \bal
 \sum_{m=1}^M r_{jm}\,\delta_{a^*}  x_m
 -\sum_{\alpha=1}^{N} {c}^{\alpha}_j  \,   \delta_{a^*} \mu_{\alpha}=0, \label{cond2-1}
 \eal
 and 
 \bal
\sum_{m=1}^M ({\mbf d}_{a})_m\,  \delta_{a^*}  x_m= \delta l_{a^*}{\color{black}\delta_{a,a*}}\label{cond2-2}.
\eal
From Eq. \eqref{cond2-1} and \eqref{cond2-2}, we obtain the matrix equation, 
 \begin{eqnarray}
{\bf A}\left(
\begin{array}{c}
\delta_{a^*} {  \mbf x} \\ \hline
\delta_{a^*} {\mbf \mu}
\end{array}
\right)\begin{array}{c }
{\updownarrow}\ _M \\
{\updownarrow}\ _{N}\\
\end{array}
= - \left(
\begin{array}{c}
{\bf 0}\\\hline
{ {\mbf e}_{a^*} }
\end{array}
\right)\begin{array}{c }
{\updownarrow}\ _R \\
{\updownarrow}\ _{N_c}\\
\end{array},\label{appcoker}
\end{eqnarray}
 where $\bf A$ is the same as that in Eq. \eqref{Amat}, and the  column vector ${\mbf e}_{a^*}$ is defined as  
$ {\mbf e}_{a^*} \equiv (0,\ldots,\underset{{a^*}{\rm{\mathchar`-}th}}{\underbrace{\delta l_{a^*}}} ,\ldots,0 )^T\in{\mathbb R}^{N_c}$. 
 
If we write the results of the perturbations for $j^*=1,\ldots, R$ and $\a = 1,\ldots, N_c$, given by Eq. \eqref{appwocoker} and \eqref{appcoker}, we obtain Eq. \eqref{dx=A-1E} in the main text.

\section{Proof of the law of localization}\label{app:proof}
Firstly, we write the precise definitions of $N({\mathfrak r})$ and $N_c({\mathfrak m})$ appearing in Eq. \eqref{generalLoL}.  For a  chemical subset ${\mathfrak m}$ and a reaction subset ${\mathfrak r}$, we can respectively associate the following vector spaces $V({\mathfrak r})$ and $V_c({\mathfrak m})$,
\begin{eqnarray}
V({\mathfrak r}) &\equiv&   \, {\rm span}\, \bigl\{  {\mbf   v}   | \,{\mbf  v} \in {\rm ker}\, \nu , P^{\mathfrak r}{\mbf v} = {\mbf v}\bigr\},  \nonumber \\ 
V_c({\mathfrak m}) &\equiv&  \, {\rm span}\, \bigl\{  P^{\mathfrak m}  {\mbf u }  | \,  {\mbf u }\in {\rm coker} \, \nu   \bigr\}.\label{vkvc}
\end{eqnarray}
Here, $P^{\mathfrak r}$ is an $R\times R$ projection matrix onto the space associated with ${\mathfrak r}$ defined as 
$$
P^{\mathfrak r}_{j,j'} = \delta_{j,j'}\   {\rm if } \  j,j' \in {\mathfrak r}.\  {\rm Otherwise } \ P^{\mathfrak r}_{j,j'} =0.
$$
In other words, $V({\mathfrak r})$ are vectors ${\mbf v}\in {\mathbb R}^R$ with component support in $\mathfrak r$. Similarly, $P^{\mathfrak m}$ is an $M\times M$ projection matrix on the space associated with ${\mathfrak m}$.  
Then, we define $N({\mathfrak r})$ and $N_c({\mathfrak m})$ as the dimensions of these vector spaces;
\begin{eqnarray}
N({\mathfrak r}) \equiv {\rm dim} \, V({\mathfrak r}), \ 
N_c({\mathfrak m})\equiv {\rm dim}\, V_c ({\mathfrak m})\label{nknc}.
\end{eqnarray}
The intuitive meaning for this definition is  explained in the main text. 
Note that $N({\mathfrak r} )={\rm dim\, ker} (\nu P^{\mathfrak r})-R+|\mathfrak r|$ and $N_c({\mathfrak m} ) = {\rm dim\, coker} \, \nu  -  {\rm dim\, coker} ( P^{\bar{\mathfrak m}}\nu )+|\mathfrak m|$, where $P^{\bar{ \mathfrak m} }\equiv \hat{\bf1}_{M\times M}- P^{ \mathfrak m}$ is a projection matrix on the space associated with the complementary  chemicals $ \bar{\mathfrak m}$ of ${\mathfrak m}$.

Now we prove the theorem.  Suppose that $\Gamma=({\mathfrak m},{ \mathfrak r})$ is a buffering structure, namely an output-complete subnetwork satisfying $\lambda(\Gamma)=0$. As discussed below, by choosing appropriate bases of the kernel and the cokernel of $\nu$ and arranging the orders of the column  and row indices of the matrix  ${\bf A}$, we can always rewrite  $\bf A$ into the form, 
\bal
{\bf A}= \begin{array}{c }
\ _{ |{\mathfrak r}|+N_{c}({\mathfrak m})}\Bigg{\updownarrow} \\
\\ 
\\
\end{array}
\overset{\overset{ |{\mathfrak m}|+N({\mathfrak r}) }{\xlongleftrightarrow[]{\hspace{3em}}}\quad \quad \quad \quad }{\left(
\begin{array}{ccc|c }
&&&  \\
&\underset{square}{\mbox{\smash{\Large $*$}}}&&\ \ \ \ \mbox{\smash{\Large $*$}}\ \ \  \\
&&& \\ \hline
&&  &\\ 
&\mbox{\smash{\Large $\bf 0$}}&& \ \ \ \mbox{\smash{\Large $*$}}\ \ \  
\end{array}
\right)}.\label{Agamma}
\eal
 Since we are assuming ${\rm Det}\, {\bf A} \neq 0$, the upper-left block in Eq. \eqref{Agamma} is generally vertically long or square; i.e. $\lambda(\Gamma)\geq 0$. The condition $\lambda(\Gamma)=0$  means that it is  square.

The structure of block matrices in Eq. \eqref{Agamma} can be obtained by collecting the indices associated with $\Gamma=({\mathfrak m},{ \mathfrak r})$ into the upper-left corner:  The column indices at the upper left block consist of the  chemicals in $ {\mathfrak m}$ followed by the basis vectors of the kernel space $V({\mathfrak r} )$ associated with $\mathfrak r$. The row indices consist of the reactions in $\mathfrak r$ followed by the basis vectors of the cokernel space $V_c({\mathfrak m})$ associated with ${\mathfrak m}$.  Thus, from this construction, the upper-left block has the size $(|{\mathfrak r}|+N_c({\mathfrak m})) \times (|{\mathfrak m}|+N({\mathfrak r}))$. We need to prove that the lower-left block vanishes completely. First,  all $r_{jm}$ with $j \notin {{ \mathfrak r}}$ and $m \in {\mathfrak m}$, which would appear in the lower-left block, vanish by the assumption that the subnetwork $\Gamma$ is output-complete; reaction rates $r_j(k_j,x)$ with $j\notin {\mathfrak r}$ do not depend on the concentration of  chemicals in  $\mathfrak m$. 
It remains to show that ${\mbf c}^\alpha \in V({\mathfrak r})$ and ${\mbf d}_a \in V_c(\mathfrak m)$ do not have nonzero entries in the lower-left block. But this directly follows from their definitions, Eq. \eqref{vkvc}. This completes the proof of the structure given in Eq. \eqref{Agamma}.



After arranging the matrix $\bf A$ as in  Eq. \eqref{Agamma}, it is  easy to prove the law of localization.
We first prove the theorem for the reaction rate perturbations, i.e.  $\delta_\j { x}_m=0$ and  $\delta_\j r_{j'}=0$ for  $\j \in {{ \mathfrak r}} \not \ni j' $, and $m \notin { {\mathfrak m}}$. As explained in the body of the paper, the concentration change $\delta_\j { x}_m$  is proportional to $({\bf A}^{-1})_{m\j}\propto \ {\rm Det}\, {\hat {\bf A}}^{\j,m}$, where ${\hat {\bf A}}^{\j,m}$ is  the minor matrix obtained by removing   reaction row $\j$  and  chemical column $m$  from the matrix $\bf A$. Noting that $\j \in {{ \mathfrak r}}$ belongs to the raw indices of the upper  part of $\bf A$, and 
$m \notin { {\mathfrak m}}$ belongs to the column indices of of the right part of $\bf A$, 
 ${\rm Det}\, {\hat {\bf A}}^{\j,m}=0$ holds for $\j \in {{ \mathfrak r}}$ and $m \notin { {\mathfrak m}}$   because the nonzero block at the upper left of the minor ${\hat {\bf A}}^{\j,m}$, which was originally square  in Eq. \eqref{Agamma},  now becomes horizontally long.  This proves $\delta_\j { x}_m=0$ for $\j \in {{ \mathfrak r}}$ and $m\notin { {\mathfrak m}}$. 
It remains to show  $\delta_\j r_{j'}=0$ for all  $\j \in {{ \mathfrak r}} \not\ni j'$. Noting  $r_{j'}$  depends on the outside chemicals $x_m$ with $m\notin \mathfrak m$ because of  the output-completeness of $\Gamma=({\mathfrak m},{\mathfrak r})$,  Eq. \eqref{dw2} becomes $\delta_\j r_{j'}=\sum_{m\notin{ \mathfrak m}} \frac{\partial r_{j'}}{\partial x_m} \delta_{j*} x_m$. Then, the chemical insensitivity 
$\delta_\j x_m =0$ for $m \not\in \mathfrak m$ also means the flux insensitivity $\delta_\j r_{j'}=0$.

{\color{black} Similarly, for the perturbations of conserved quantities,  we can prove the chemical insensitivity, $\delta_{a^*} x_m=0$ for  all $m \notin \Gamma$ and the flux insensitivity  $\delta_{a^*}  r_{j'}=0$ for all  $j' \notin {{ \mathfrak r}}$, under any perturbation of conserved quantities in $\Gamma$, that is, any perturbation of $l_{a^*} ={\mbf d}_{a^*}\cdot {\mbf x}$  for  ${\mbf d}_{a^*} \in V_c({\mathfrak m})$. From \eqref{appcoker}, $\delta_{a^*} x_m$ is proportional to  $({\bf A}^{-1})_{m a^*}\propto \ {\rm Det}\, {\hat {\bf A}}^{a^*,m}$, the determinant of the minor matrix obtained by removing  the row of the $a^*$-th conserved quantity. Noting that  $m \notin \Gamma$ belongs to the column indices of the right part of $\bf A$ and  the conserved quantity $l_{a^*}$  associated with ${\mbf d}_{a^*} \in V_c({\mathfrak m})$ belongs to the row indices of the upper part of $\bf A$, we can show ${\rm Det}\, {\hat {\bf A}}^{a^*,m} =0$ and $\delta_{a^*} x_m=0$, which leads to the flux insensitivity  $\delta_{a^*}  r_{j'}=0$, as in the above argument for the reaction rate perturbations. 
}
 This proves the law of localization. \hspace{\fill}$\square$

\section{Example network}
\begin{figure}[h] 
  \includegraphics[width=6cm,bb=0 20 400 250]{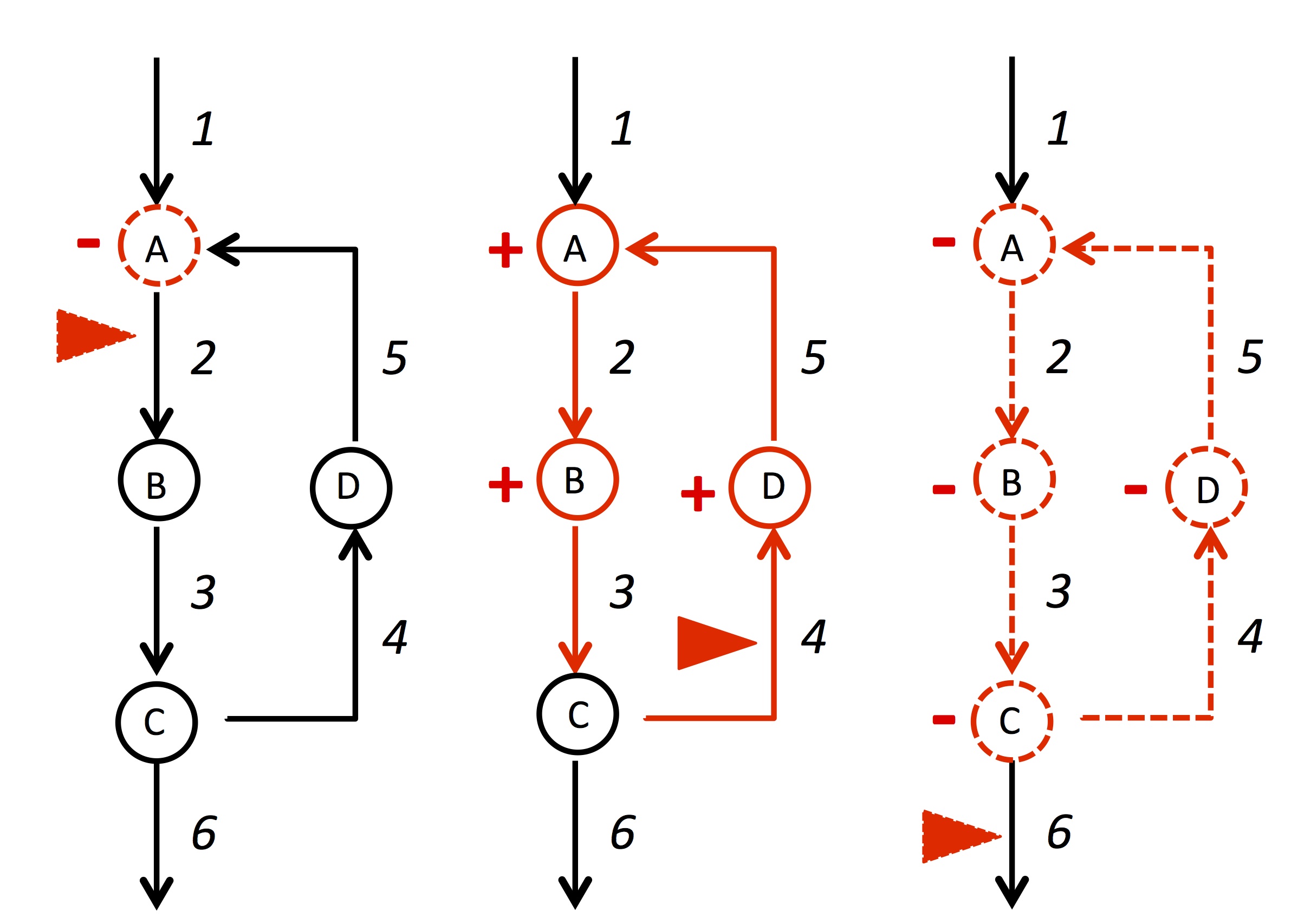}
   \caption{ Reaction networks and sensitivities in Example 1 and 2. Red triangle indicate  overexpressed reactions. The signs (increase/decrease) of responses are represented by  $+/-$ for chemicals and solid/dashed red lines  for fluxes. }
   \label{fig:rei1}
\end{figure}

We illustrate the computation of the sensitivity analysis for the network consisting of $R=6$ reactions and $M=4$ chemicals,  shown in FIG. \ref{fig:rei1}. 

The stoichiometric matrix $\nu$ is
\bal\nu=\left(
\begin{array}{cccccc}
 1 & -1 & 0 & 0 & 1 & 0 \\
 0 & 1 & -1 & 0 & 0 & 0 \\
 0 & 0 & 1 & -1 & 0 & -1 \\
 0 & 0 & 0 & 1 & -1 & 0 \\
\end{array}
\right).
\eal
Noting that $N_c=0$,  the matrices {\bf A} and $\bf S$ are
\bal
\scalebox{0.8}{$ \A = \left(
    \begin{array}{cccc|cc}
0 & 0 & 0 & 0 & -1& 0 \\
r_{2A} & 0 & 0 & 0 & -1& -1\\
0 & r_{3B} & 0 & 0 & -1& -1\\
0 & 0 & r_{4C} & 0 & 0 & -1\\
0 & 0 & 0 & r_{5D} & 0 & -1\\
0 & 0 & r_{6C} & 0 & -1 & 0\\
\end{array}
  \right), 
  $}
\eal
\bal
\scalebox{0.8}{$ 
{\bf S} = 
\left(
    \begin{array}{cccccc}
\frac{-r_{4C}-r_{6C}}{r_{2A}r_{6C}} & r_{2A}^{-1} & 0 & - r_{2A}^{-1} & 0& \frac{r_{4C}}{r_{2A}r_{6C}} \\
\frac{-r_{4C}-r_{6C}}{r_{3B}r_{6C}} & 0 & r_{3B}^{-1} & -r_{3B}^{-1} & 0 &  \frac{r_{4C}}{r_{3B}r_{6C}} \\
- \frac{1}{r_{6C}}& 0 & 0 & 0 & 0&  r_{6C}^{-1} \\ 
 -\frac{r_{4C}}{r_{5D}r_{6C}} & 0 & 0 & -r_{5D}^{-1} &r_{5D}^{-1} &  \frac{r_{4C}}{r_{5D}r_{6C}} \\ \hline
-1 & 0 & 0 & 0 & 0& 0 \\
- \frac{r_{4C}}{r_{6C}} & 0 & 0 & -1 & 0& \frac{r_{4C}}{r_{6C}}.
\end{array}
  \right).$}\label{Sense2}
\eal
Then, from Eqs. \eqref{dx=A-1E} and \eqref{fluxresponse}, the responses of chemical concentrations and fluxes are 
\bal
\delta_\j x_m=\left(
\begin{array}{cccccc}
 \frac{r_{4 {C}}+r_{6 {C}}}{r_{2 {A}} r_{6 {C}}} & \frac{-1}{r_{2 {A}}} & 0 & \frac{1}{r_{2 {A}}} & 0 & \frac{-r_{4 {C}}}{r_{2 {A}} r_{6 {C}}} \\
 \frac{r_{4 {C}}+r_{6 {C}}}{r_{3 {B}} r_{6 {C}}} & 0 & \frac{-1}{r_{3 {B}}} & \frac{1}{r_{3 {B}}} & 0 & \frac{-r_{4 {C}}}{r_{3 {B}} r_{6 {C}}} \\
 \frac{1}{r_{6 {C}}} & 0 & 0 & 0 & 0 & \frac{1}{-r_{6 {C}}} \\
 \frac{r_{4 {C}}}{r_{5 {D}} r_{6 {C}}} & 0 & 0 & \frac{1}{r_{5 {D}}} & -\frac{1}{r_{5 {D}}} & \frac{-r_{4 {C}}}{r_{5 {D}} r_{6 {C}}} \\
\end{array}
\right)_{m\j},\label{rei1}
\eal
and
\bal
\delta_\j r_j=\left(
\begin{array}{cccccc}
 1 & 0 & 0 & 0 & 0 & 0 \\
 \frac{r_{4 {C}}+r_{6 {C}}}{r_{6 {C}}} & 0 & 0 & 1 & 0 & -\frac{r_{4 {C}}}{r_{6 {C}}} \\
 \frac{r_{4 {C}}+r_{6 {C}}}{r_{6 {C}}} & 0 & 0 & 1 & 0 & -\frac{r_{4 {C}}}{r_{6 {C}}} \\
 \frac{r_{4 {C}}}{r_{6 {C}}} & 0 & 0 & 1 & 0 & -\frac{r_{4 {C}}}{r_{6 {C}}} \\
 \frac{r_{4 {C}}}{r_{6 {C}}} & 0 & 0 & 1 & 0 & -\frac{r_{4 {C}}}{r_{6 {C}}} \\
 1 & 0 & 0 & 0 & 0 & 0 \\
\end{array}
\right)_{j\j}.\label{exampledr}
\eal
We can see that only the perturbation to the input rate, corresponding to the 1st column in Eq. \eqref{Sense2},  affects all chemicals and fluxes. The perturbations to reactions $2, 3, 5$ only decrease the concentrations of the substrates $A, B, D$ respectively. 
The perturbation of reaction $4$ decreases the  concentrations $D, A, B$ along the  cycle downward of the perturbation (see FIG. \ref{fig:rei1}, and the 4th column of $\bf S$). The perturbation of reaction 6 does not change  the further downstream but change $A,B,C,D$ in the  cycle. 

\begin{figure}[htbp] 
  \includegraphics[width=6cm,bb=0 0 400 300]{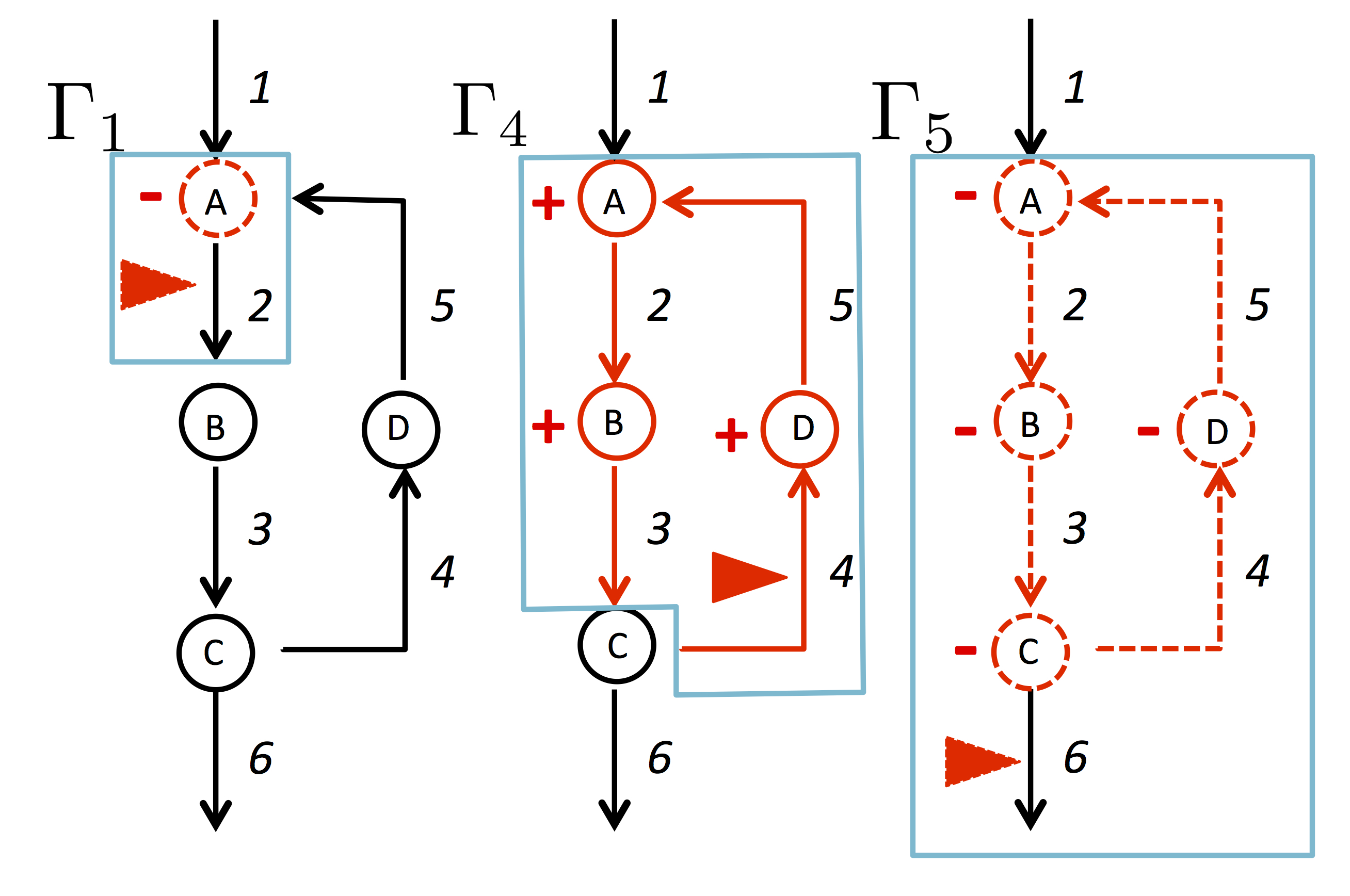}
   \caption{The boxes represent the buffering structures, $\Gamma_2,\ \Gamma_4$ and $\Gamma_5$.}
   \label{fig:bs_rei1}
\end{figure}

The law of localization can be applied as follows. Some of the buffering structures are shown in FIG. \ref{fig:bs_rei1}. The smallest buffering structures are $\G_1=(\{A\},\{2\}),\  \G_2=(\{B\},\{3\}),\ \G_3=(\{D\},\{5\})$, which all sassily $\lambda(\Gamma_i)=-1+1-0=0$. In addition, the network has two larger ones, $\G_4=(\{A,B,D\},\{2,3,4,5\})$ (with $\lambda(\G_4)=-3+4-1=0$), $\G_5=(\{A,B,C,D\},\{2,3,4,5,6\})$ (with $\lambda(\G_5)=-4+5-1=0$). $\G_4$ is the minimum buffering structure including reaction 4. Then,  the {\it law of localization} predicts that the nonzero response to perturbation of reaction 4 should be limited within $\G_4$, which is observed in the 4th column in Eq. \eqref{Sense2}. Similarly, the response to perturbation of reaction 6 is explained by $\G_5$. 

\section{ E. coli  central metabolism}\label{app:ecoli}
\subsubsection{{ List of reactions}}\ 
1: Glucose  +  PEP  $\rightarrow$  G6P  +  PYR. 
 
 2: G6P   $\leftarrow$  F6P. 
 
 3: F6P   $\rightarrow$  G6P.
  
  4: F6P  $\rightarrow$  F1,6P. 
   
   5: F1,6P  $\rightarrow$  G3P  +  DHAP. 
   
   6: DHAP   $\rightarrow$  G3P.
   
 7: G3P   $\rightarrow$  3PG. 
  
  8: 3PG   $\rightarrow$  PEP. 
  
  9: PEP   $\rightarrow$  3PG.  
  
  10: PEP   $\rightarrow$  PYR. 
  
  11: PYR   $\rightarrow$  PEP. 
  
  12: PYR    $\rightarrow$  AcCoA  +   CO2. 
  
  13: G6P  $\rightarrow$  6PG. 
  
  14: 6PG  $\rightarrow$   Ru5P  +  CO2. 
  
  15: Ru5P  $\rightarrow$  X5P.
  
   16: Ru5P  $\rightarrow$   R5P. 
   
   17: X5P  +  R5P  $\rightarrow$   G3P  +  S7P. 
   
   18: G3P  +  S7P  $\rightarrow$   X5P  +  R5P. 
   
   19: G3P  +  S7P  $\rightarrow$   F6P  +  E4P.
   
    20: F6P  +  E4P  $\rightarrow$  G3P  +  S7P. 
    
21:  X5P  +  E4P  $\rightarrow$   F6P  +  G3P. 
  
  22: F6P   +  G3P  $\rightarrow$   X5P  +  E4P.  
  
  23: AcCoA  +    $\rightarrow$  CIT. 
  
  24: CIT   $\rightarrow$  ICT. 
  
  25: ICT  $\rightarrow$  2${\rm \mathchar`-}$KG  +  CO2. 
  
  26: 2-KG  $\rightarrow$   SUC  +  CO2.
  
   27: SUC   $\rightarrow$  FUM. 
   
  28:  FUM  $\rightarrow$  MAL. 
   
   29: MAL   $\rightarrow$  OAA.
   
 30: OAA   $\rightarrow$  MAL.

 31: PEP  +  CO2  $\rightarrow$  OAA.

 32: OAA  $\rightarrow$  PEP  +   CO2. 

 33: MAL  $\rightarrow$   PYR  +  CO2.

34: ICT   $\rightarrow$  SUC  +  Glyoxylate. 

 35: Glyoxylate  +  AcCoA  $\rightarrow$  MAL. 

 36: 6PG  $\rightarrow$   G3P  +  PYR. 

 37: AcCoA  $\rightarrow$   Acetate. 

38:  PYR  $\rightarrow$  Lactate. 

 39: AcCoA  $\rightarrow$  Ethanol. 

 40: R5P  $\rightarrow$ (output).

 41: OAA  $\rightarrow$ (output).

 42: CO2  $\rightarrow$ (output).

43:  (input) $\rightarrow$  Glucose. 

 44:  Acetate $\rightarrow$ (output).
 
  45: Lactate $\rightarrow$ (output).

46:  Ethanol $\rightarrow$ (output).

\subsubsection{{ List of buffering structures}} \
The E. coli network exhibits the following 17 different buffering structures $\G_i=(\m_i,\r_i) $  ($i=1,\ldots,17$).

{\small 
$\Gamma_1=(\{ \rm 
Glucose
\},\{
1
\})$,

$\Gamma_2=(\{ \rm 
Glucose,
PEP,
G6P,
F6P,
F1,6P,
DHAP,
G3P,
3PG,\\
PYR,
6PG,
Ru5P,
X5P,
R5P, 
S7P,
E4P,
AcCoA,
OAA,
CIT,\\
ICT,
2{\rm \mathchar`-KG},
SUC,
FUM,
MAL,
CO2,
Glyoxylate,
Acetate,\\
Lactate,
Ethanol
\},  \{
1,
2 ,
3 ,
4 ,
5 ,
6 ,
7 ,
8 ,
9 ,
10 ,
11 ,
12 ,
13 ,
14 ,
15 ,\\
16 ,
17 ,
18 ,
19 ,
20 ,
21 ,
22 ,
23 ,
24 ,
25 ,
26 ,
27 , 
28 ,
29 ,
30 ,
31 ,
32 ,
33 ,\\
34 ,
35 ,
36 ,
37 ,
38 ,
39 ,
40 ,
41 ,
42 ,
44 ,
45 ,
46 
\})$,

$\Gamma_3=(\{ \rm 
F1,6P
\},\{
5 
\})$,

$\Gamma_4=(\{ \rm 
DHAP
\},\{
6 
\})$,

$\Gamma_5=(\{ \rm 
G3P,
X5P,
S7P,
E4P
\},\{
7 ,
17 ,
18 ,
19 ,
20 ,
21 ,
22 
\})$,

$\Gamma_6=(\{ \rm 
3PG
\},\{
8 
\})$,

$\Gamma_7=(\{ \rm 
Glucose,
PEP,
3PG,
PYR,
AcCoA,
OAA,
CIT,
ICT,\\
2{\rm \mathchar`-KG},
SUC,
FUM, 
MAL,
CO2,
Glyoxylate,
Acetate,
Lactate,\\
Ethanol
\},\{
1,
8 ,
9 ,
10 ,
11 ,
12 ,
23 ,
24 ,
25 ,
26 ,
27 ,
28 ,
29 ,
30 ,
31 ,\\
32 ,
33 ,
34 ,
35 ,
37 ,
38 ,
39 ,
41 ,
42 ,
44 ,
45 ,
46 
\})$,

$\Gamma_8=(\{ \rm 
X5P,
S7P,
E4P
\},\{
17 ,
18 ,
19 ,
20 ,
21 
\})$ ,

$\Gamma_9=(\{ \rm 
CIT
\},\{
24 
\})$,

$\Gamma_{10}=(\{ \rm 
2{\rm \mathchar`-KG}
\},\{
26 
\})$,

$\Gamma_{11}=(\{ \rm 
SUC
\},\{
27 
\})$ ,

$\Gamma_{12}=(\{ \rm 
FUM
\},\{
28 
\})$ 

$\Gamma_{13}=(\{ \rm 
Glyoxylate
\},\{
35 
\})$,

$\Gamma_{14}=(\{ \rm 
X5P,
R5P,
S7P,
E4P
\},\{
17 ,
18 ,
19 ,
20 ,
21 ,
40 
\})$ ,

$\Gamma_{15}=(\{ \rm 
Acetate
\},\{
44 
\})$,

$\Gamma_{16}=(\{ \rm 
Lactate
\},\{
45 
\})$,

$\Gamma_{17}=(\{ \rm 
Ethanol
\},\{
46 
\})$.
}

\section{The degrees and the values of robustness in rewired E. coli networks}

\begin{figure}[htbp]
  \includegraphics[width=6cm,bb=0 0 330 220]{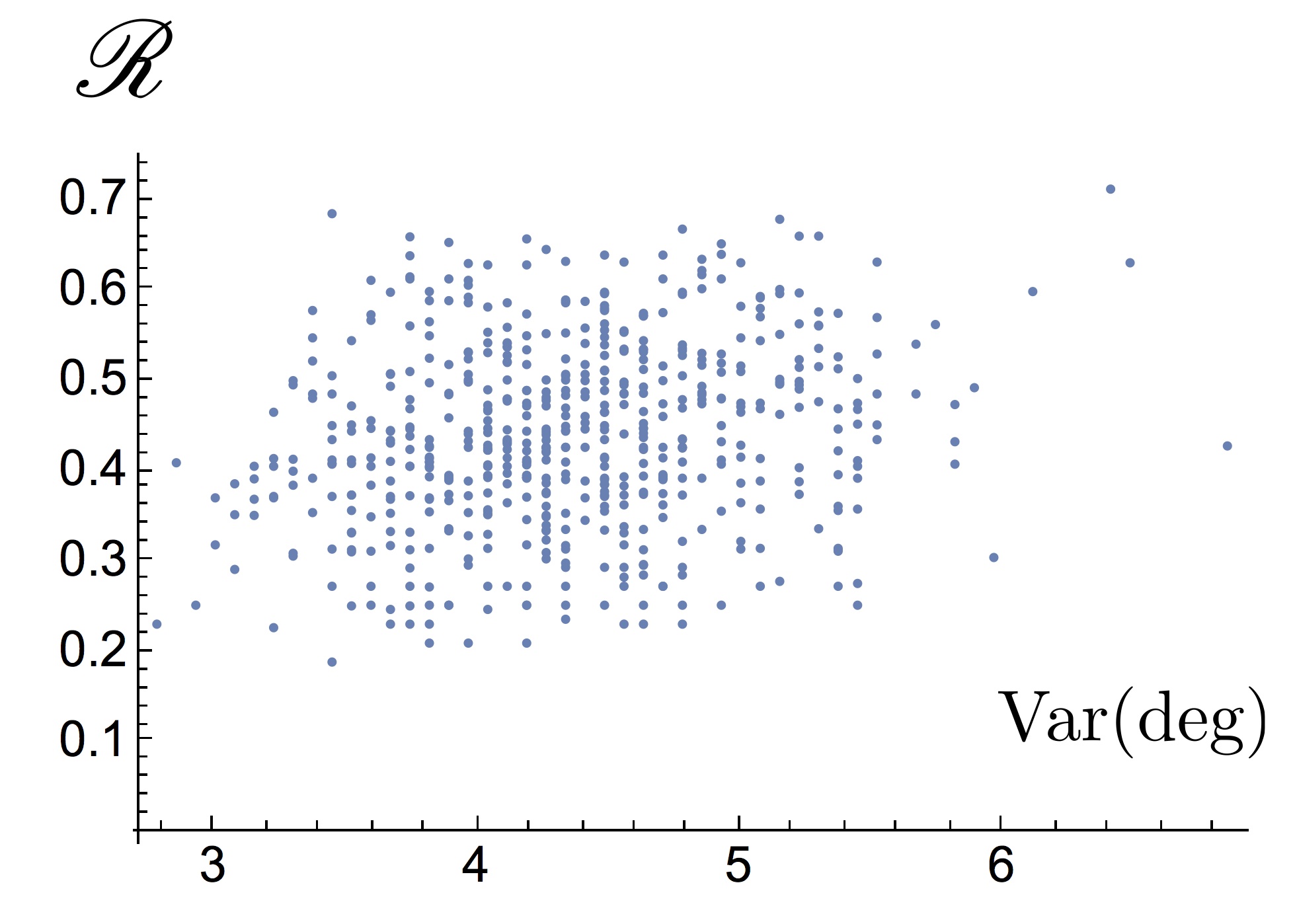}
   \caption{The distribution of the  rewired networks when $p=4$}  
   \label{fig:vertex}
\end{figure}

{\color{black} We define the degree  of the $m$-th metabolite as the number of reactions with which the  $m$-th metabolite participates is involved as a substrate or product. 
Here, we computed the variance of degrees and  $\mathscr R$ for each rewired network, and investigated whether there exist any correlation between these two quantities. We note that the average of degrees in a network are the same for all rewired networks   because the rewiring procedure preserves the total number of reactions. 
FIG. \ref{fig:vertex} shows the distributions of  the variance of degree and $\mathscr R$ for 556 rewired networks when $p=4$. We did not observe any strong correlation between them, as we mentioned in the main text. }

\section{The Watts-Strogatz-like model}
Here, we consider  a model of random reaction networks similar to the Watts-Strogatz model \cite{WS}.
\begin{figure}[htbp]
  \includegraphics[width=7cm,bb=0 0 180 100]{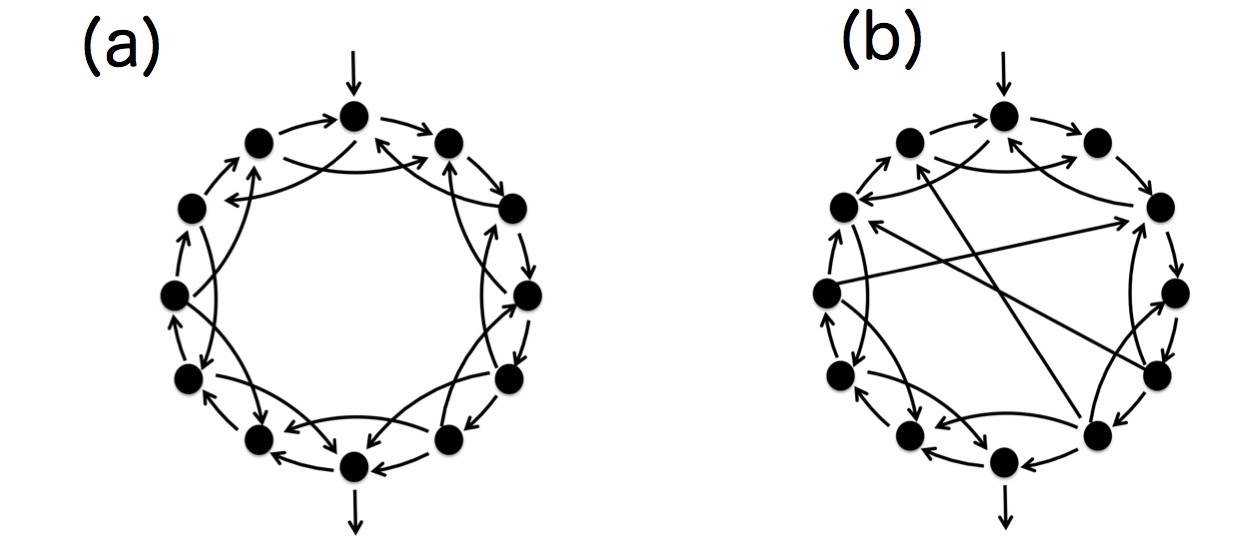}
   \caption{(a) The Watts-Strogatz-like reaction system with $M=12$ and $p=0$. The directions of the internal reactions are randomly chosen. (b) An example of randomly rewired networks, where the internal edges are randomly rewired with probability $p$. The reactions along the circle are fixed.  }  
   \label{fig:wattsstrogatz}
\end{figure}
\begin{figure}[htbp]
  \includegraphics[width=12cm,bb=0 0 600 240]{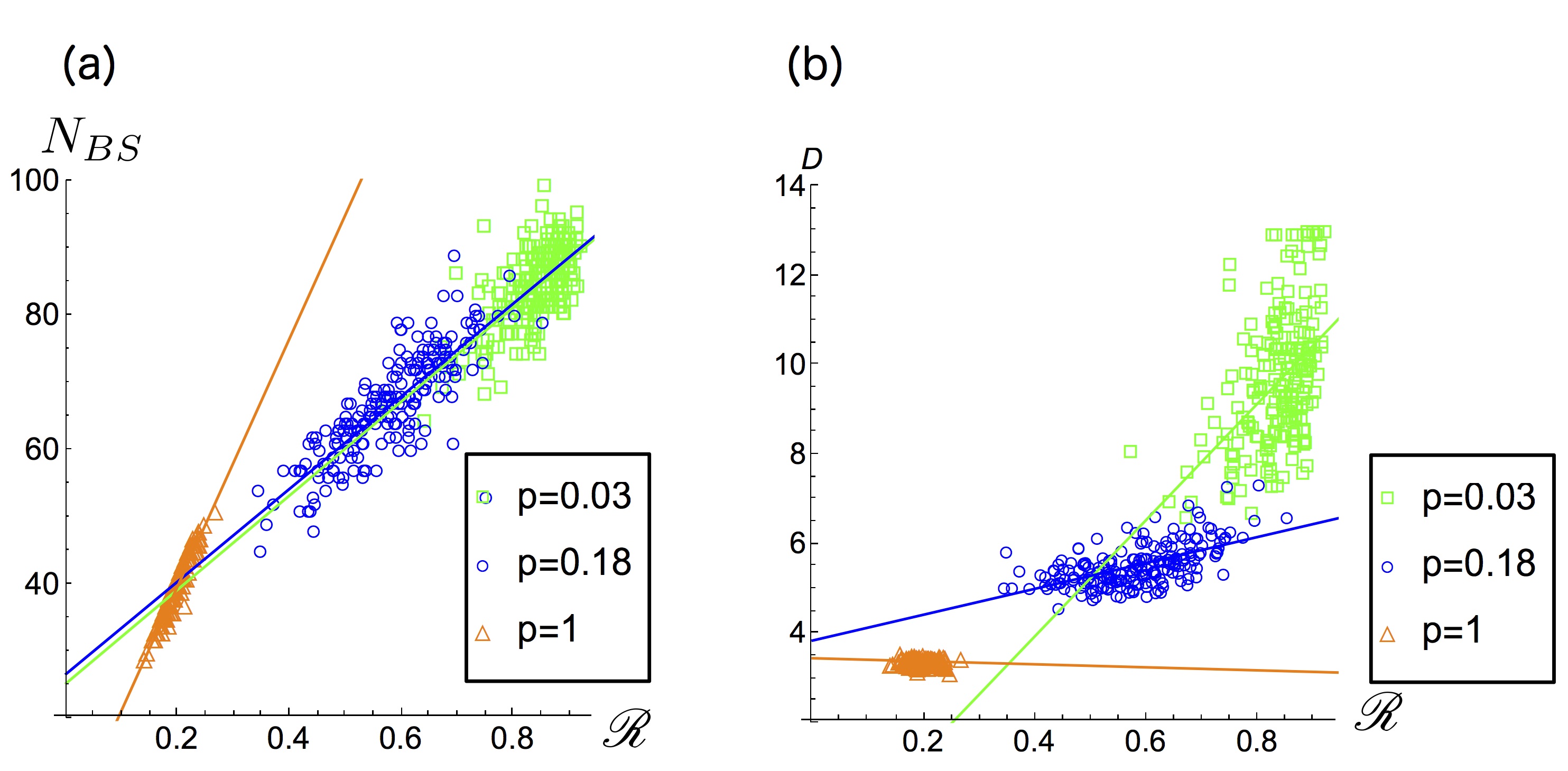}
  \caption{ (a)The distribution of random networks in ${\mathscr R}-\ N_{BS}$ plane.  Pearson correlation coefficient (averaged over network ensembles labeled by $p$) is 0.77. (b) The distribution of random networks in ${\mathscr R}-D$ plane. Pearson correlation coefficient (averaged over network ensembles labeled by $p$) is 0.36. }
 \label{fig:wshistogram}
\end{figure}
\begin{figure}[htbp]
  \includegraphics[width=8cm,bb=-50 0 300 200]{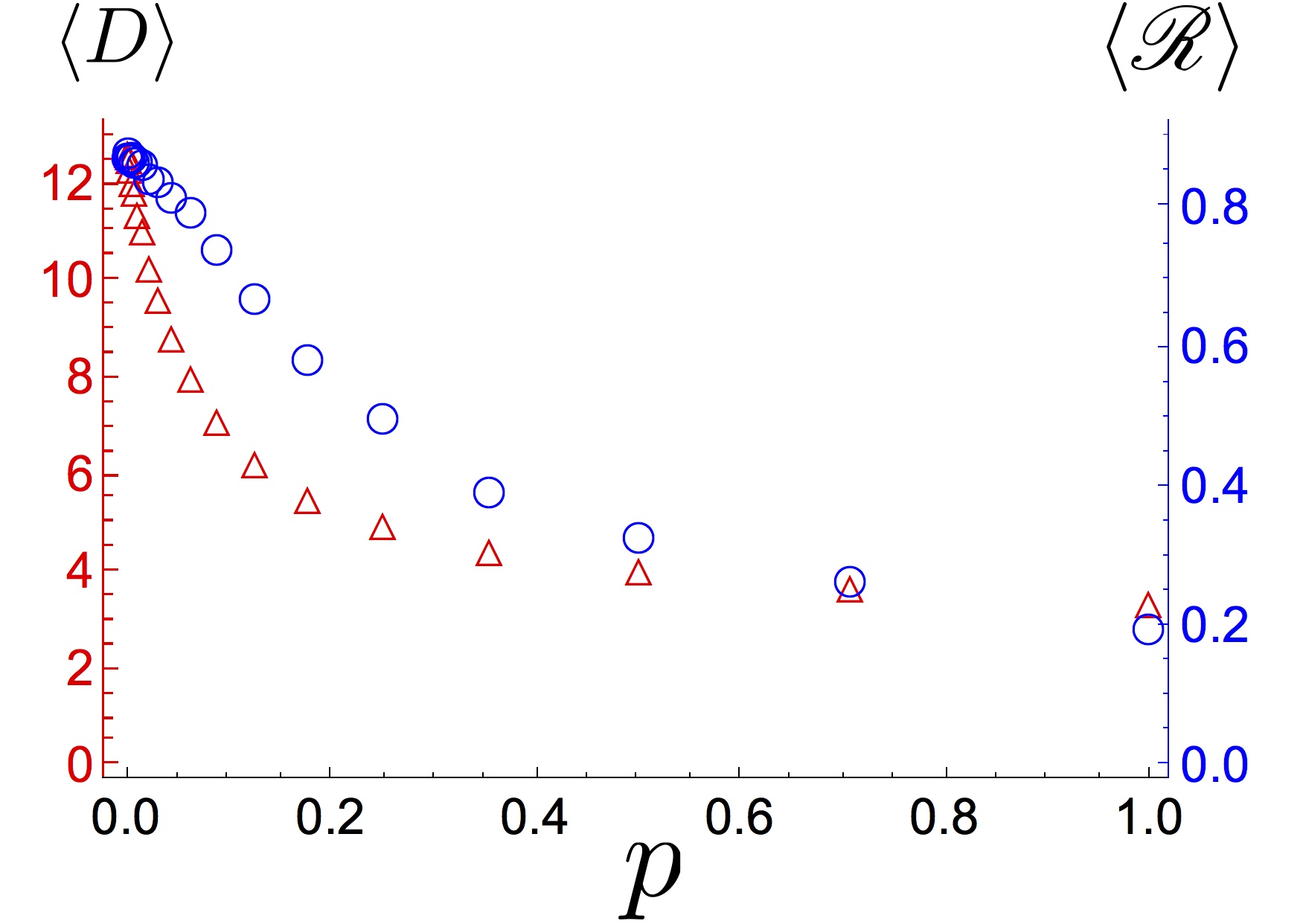}
   \caption{ The averages of ${\mathscr R}$ and $D$ over the network ensemble with fixed $p$ ($0 \leq p \leq1$).}
   \label{fig:resultWS}
\end{figure}
The model consists  of  $M$ chemicals and reactions along a circular clockwise pathway and randomly directed reactions inside the circle.  The  end points of internal reactions are randomly rewired from the next neighbors to other chemicals with probability $p$. There is  an inflow and an outflow on the circle.  In fact, $ {\mathscr R} $ and $ N_{BS} $ are independent of the positions of the inflow and the outflow. We note that, unlike the original Watts-Strogatz model, the reactions along the circle are not rewired, which guarantees  the existence of the inverse of $\bf A$ in Eq. (11). 

We generated 1000 random networks for each ensemble labeled by $p$.
Two figures in FIG. \ref{fig:wshistogram} show the distributions of  ${\mathscr R},N_{BS}$ and of ${\mathscr R},D$. We can see that while $N_{BS}$ and ${\mathscr R}$ are all positively correlated for any value of $p$, ${\mathscr R}$ and $D$ are not correlated significantly.   

FIG. \ref{fig:resultWS} shows the result of $\langle {\mathscr R} \rangle $ and $\langle D \rangle $ for various $p$.  As we randomize the network by increasing $p$, the mean distance  $\langle D \rangle $ and  $\langle {\mathscr R} \rangle $ decrease monotonically.  In contrast to the E. coli network, there is no strong peak of the robustness ${\mathscr R}$ around $p=0$ in the Watts-Strogatz-like model.

\nocite{*}


\end{document}